\documentclass[trackchanges,twocolumn,twocolappendix]{aastex701}

\usepackage{amsmath}
\usepackage{multirow}
\usepackage{float}

\newcommand{\Lya}{Ly$\alpha$ }

\begin{document}

\title{ODIN: Rest-frame Optical Morphologies and Star Formation Activity of \Lya Emitters at $\bf{z=2.4}$, $\bf{3.1}$, and $\bf{4.5}$}

\author[orcid=0009-0003-9748-4194]{Sang Hyeok Im}
\affiliation{Department of Physics and Astronomy, Seoul National University, 1 Gwanak-ro, Gwanak-gu, Seoul 08826, Republic of Korea}
\email[show]{sanghyeok.im97@gmail.com}  
\affiliation{Korea Institute for Advanced Study, 85 Hoegi-ro, Dongdaemun-gu, Seoul 02455, Republic of Korea}

\author[orcid=0000-0003-3428-7612]{Ho Seong Hwang} 
\affiliation{Department of Physics and Astronomy, Seoul National University, 1 Gwanak-ro, Gwanak-gu, Seoul 08826, Republic of Korea}
\affiliation{SNU Astronomy Research Center, Seoul National University, 1 Gwanak-ro, Gwanak-gu, Seoul 08826, Republic of Korea}
\affiliation{Institute for Data Innovation in Science, Seoul National University, Seoul 08826, Korea}
\email[show]{hhwang@astro.snu.ac.kr}

\author[orcid=0000-0003-3301-759X]{Jeong Hwan Lee} 
\affiliation{Department of Physics and Astronomy, Seoul National University, 1 Gwanak-ro, Gwanak-gu, Seoul 08826, Republic of Korea}
\affiliation{Research Institute of Basic Sciences, Seoul National University, Seoul 08826, Republic of Korea}
\email{joungh93@gmail.com}

\author[orcid=0000-0002-1328-0211]{Robin Ciardullo} 
\affiliation{Concha 700, Las Condes, Santiago RM, Chile}
\affiliation{Institute for Gravitation and the Cosmos, The Pennsylvania State University, University Park, PA 16802}
\email{}

\author[orcid=0000-0003-1530-8713]{Eric Gawiser} 
\affiliation{Department of Physics and Astronomy, Rutgers, the State University of New Jersey, Piscataway, NJ 08854, USA}
\email{}

\author[orcid=0000-0001-6842-2371]{Caryl Gronwall} 
\affiliation{Department of Astronomy \& Astrophysics, The Pennsylvania State University, University Park, PA 16802, USA}
\affiliation{Institute for Gravitation and the Cosmos, The Pennsylvania State University, University Park, PA 16802, USA}
\email{}

\author[orcid=0000-0002-4902-0075]{Lucia Guaita} 
\affiliation{Universidad Andres Bello, Facultad de Ciencias Exactas, Departamento de Fisica y Astronomia, Instituto de Astrofisica, Fernandez Concha 700, Las Condes, Santiago RM, Chile}
\affiliation{Millennium Nucleus for Galaxies (MINGAL), Chile}
\email{}

\author[orcid=0000-0002-2770-808X]{Woong-Seob Jeong} 
\affiliation{Korea Astronomy and Space Science Institute, 776, Daedeokdae-ro, Yuseong-gu, Daejeon 34055, Republic of Korea}
\email{}

\author[orcid=0000-0001-6270-3527]{Ankit Kumar} 
\affiliation{Universidad Andres Bello, Facultad de Ciencias Exactas, Departamento de Fisica y Astronomia, Instituto de Astrofisica, Fernandez}
\email{}

\author[orcid=0000-0003-3004-9596]{Kyoung-Soo Lee} 
\affiliation{Department of Physics and Astronomy, Purdue University, 525 Northwestern Avenue, West Lafayette, IN 47906, USA}
\email{}

\author[orcid=0000-0001-9521-6397]{Changbom Park} 
\affiliation{Korea Institute for Advanced Study, 85 Hoegi-ro, Dongdaemun-gu, Seoul 02455, Republic of Korea}
\email{}

\author[orcid=0000-0002-9176-7252]{Vandana Ramakrishnan} 
\affiliation{Department of Physics and Astronomy, Purdue University, 525 Northwestern Avenue, West Lafayette, IN 47907, USA}
\email{}

\author[orcid=0009-0000-7651-3900]{Akriti Singh} 
\affiliation{Universidad Andres Bello, Facultad de Ciencias Exactas, Departamento de Fisica y Astronomia, Instituto de Astrofisica, Fernandez Concha 700, Las Condes, Santiago RM, Chile}
\email{}

\author[orcid=0000-0002-4362-4070]{Hyunmi Song} 
\affiliation{Department of Astronomy and Space Science and Institute for Sciences of the Universe, Chungnam National University, Daejeon 34134, Republic of Korea}
\email{}

\author[orcid=0000-0001-9991-8222]{Sungryong Hong} 
\affiliation{Korea Astronomy and Space Science Institute, 776 Daedeokdae-ro, Yuseong-gu, Daejeon, Korea 34055}
\email{}

\author[orcid=0000-0002-4391-2275]{Juhan Kim} 
\affiliation{Center for Advanced Computation, Korea Institute for Advanced Study, 85 Hoegiro, Dongdaemun-gu, Seoul 02455, Republic of Korea}
\email{}

\author[orcid=0000-0002-6810-1778]{Jaehyun Lee} 
\affiliation{Korea Astronomy and Space Science Institute, 776, Daedeokdae-ro, Yuseong-gu, Daejeon 34055, Republic of Korea}
\email{}

\author[orcid=0000-0003-0695-6735]{Christophe Pichon} 
\affiliation{Korea Institute for Advanced Study, 85 Hoegi-ro, Dongdaemun-gu, Seoul 02455, Republic of Korea}
\affiliation{CNRS and Sorbonne Université, UMR 7095, Institut d’Astrophysique de Paris, 98 bis, Boulevard Arago, F-75014 Paris, France}
\affiliation{Kyung Hee University, Dept. of Astronomy \& Space Science, Yongin-shi, Gyeonggi-do 17104, Republic of Korea}
\email{}

\author[orcid=0000-0002-0930-6466]{Caitlin M. Casey} 
\affiliation{Department of Physics, University of California, Santa Barbara, CA 93106, USA}
\email{}

\author[orcid=0000-0002-3560-8599]{Maximilien Franco} 
\affiliation{Université Paris-Saclay, Université Paris Cité, CEA, CNRS, AIM, 91191 Gif-sur-Yvette, France}
\email{}

\author[orcid=0000-0003-0129-2079]{Santosh Harish} 
\affiliation{Laboratory for Multiwavelength Astrophysics, School of Physics and Astronomy, Rochester Institute of Technology, 84 Lomb Memorial Drive, Rochester, NY 14623, USA}
\email{}

\author[orcid=0000-0001-9187-3605]{Jeyhan S. Kartaltepe} 
\affiliation{Laboratory for Multiwavelength Astrophysics, School of Physics and Astronomy, Rochester Institute of Technology, 84 Lomb Memorial Drive, Rochester, NY 14623, USA}
\email{}

\begin{abstract}

We analyze the rest-frame optical ($\sim8000\;\rm{\AA}$) morphologies and star formation activity of \Lya emitters (LAEs) at redshifts $2.4$, $3.1$, and $4.5$, identified in the ODIN survey. To compare their physical properties with those of other galaxies, we construct a comparison sample of typical star-forming galaxies (SFGs) at similar redshifts from the COSMOS2025 catalog. Using the \textit{JWST}/NIRCam images from the COSMOS-Web survey, we measure the rest-frame optical sizes and Sérsic indices. We first examine their size-mass relations and find that LAEs at all three redshifts have smaller sizes than typical SFGs, with the size difference decreasing at higher redshifts. We also find that LAEs tend to have larger Sérsic indices at $z=2.4$ and $3.1$ than typical SFGs, but the difference becomes weaker at $z=4.5$. These trends are qualitatively reproduced in the Horizon Run 5 cosmological hydrodynamical simulation. We then investigate star formation activity and find that LAEs exhibit higher star formation rates than typical SFGs at all redshifts considered. Finally, we examine the connection between \Lya emission and galaxy structure, finding that the rest-frame equivalent width (REW) of the \Lya emission line has negative and positive correlations with size and Sérsic index, respectively. In addition, we find a strong positive correlation between the \Lya REW and the ratio of the instantaneous star formation rate to that averaged over the last $100\;\mathrm{Myr}$ (i.e., $\mathrm{SFR_{inst}}/\mathrm{SFR_{100 Myr}}$). These results suggest the compact and starbursting nature of LAEs, and provide important constraints on the physical mechanism for the \Lya photon escape from galaxies.

\end{abstract}


\section{Introduction} 

Studying the large-scale distribution of galaxies and its impact on galaxy properties at high redshifts is crucial for understanding the formation and evolution of galaxies \citep[e.g., ][]{Shimakawa_2015_MNRAS, Laigle_2018_MNRAS, Perez-Martinez_2024_MNRAS}. Since \cite{Partridge_Peebles_1967_ApJ} presented the possibility of using the redshifted \Lya emission line to study star-forming galaxies at high redshifts, \Lya emitters (LAEs) have become a valuable population for studying large-scale structures and galaxy evolution at high redshifts \citep[e.g., ][]{Ouchi_2005_ApJ, KSLee_2014_ApJ, Im_2024_ApJ, Ramakrishnan_2024_ApJ_filaments}. LAEs are generally known as young, low-dust, and low-mass star-forming galaxies, undergoing bursts of star formation \citep[e.g., ][]{Gawiser_2007_ApJ, Guaita_2010_ApJ, Firestone_2025_ApJ_LAE_SFHs}. They are also considered as one of important contributors to the cosmic reionization \citep[e.g.,][]{Simmonds_2023_MNRAS, Jones_2025_MNRAS}.

However, the exact physical mechanisms that govern the production and escape of \Lya photons from the interstellar and circumgalactic medium remain poorly understood. This is primarily because \Lya photons undergo complex resonant scattering with neutral hydrogen or are absorbed by dust while escaping galaxies (e.g., \citealt{Neufeld_1990, Ahn_2003_MNRAS, Verhamme_2006, Hansen_2006, Verhamme_2008, Laursen_2009, Gronke_2015, Dijkstra_2017, Kakiichi_Dijkstra_2018, Smith_2019, Song_2020_ApJ, Yu_2025_ApJ,seo2026}). Therefore, it is challenging to directly connect the observed properties of the \Lya emission line (e.g., its luminosity, spectral shape, or spatial extent) to the intrinsic properties of the host galaxy.

A common approach to studying the physical conditions for \Lya photon escape is to examine the distinct features of LAEs compared to the entire population of star-forming galaxies (SFGs) or those not selected based on \Lya emission. For example, studies using the Hubble Space Telescope (\textit{HST}) find that LAEs have smaller sizes than star-forming galaxies \citep[e.g., ][]{Bond_2009_ApJ, Malhotra_2012_ApJ, Paulino-Afonso_2018_MNRAS, Shibuya_2019_ApJ, Kim_2026_ApJ}. However, such studies were limited to rest-frame UV morphologies, which are strongly affected by dust and clumpy star-forming regions within galaxies. Only recently, there are several studies using the James Webb Space Telescope \citep[\textit{JWST}; ][]{Gardner_2023}, confirming their compact nature of LAEs at rest-frame optical morphologies \citep[e.g., ][]{Liu_2024_ApJ, Ning_2024_ApJ, Song2026size}. Nevertheless, a robust comparison with SFGs remains limited by small sample sizes, broad redshift ranges, or the reliance on extrapolated scaling relations.

On the other hand, there are also numerous studies that examine the physical properties of LAEs including stellar mass and star formation activity. These properties are usually obtained through spectral energy distribution (SED) fittings, which requires robust multi-wavelength photometry. Numerous studies perform SED fittings of LAEs and find that they are low-mass, young, and starbursting populations \citep[e.g., ][]{Gawiser_2007_ApJ, Guaita_2011_ApJ, Hao_2018_ApJ, Shimizu_2025_arXiv}. However, most of these studies rely on parametric star formation history (SFH) models, which can impose priors on SFH reconstruction and bias the resulting stellar mass or star formation rate measurements \citep[e.g., ][]{Pacifici_2015_MNRAS, Iyer_2017, Carnall_2019_ApJ, Iyer_2019, Leja_2019_ApJ}.

The One-hundred-deg$^2$ DECam Imaging in Narrowband \citep[ODIN;][]{KSLee_2024_ApJ} is an ongoing survey to identify LAEs at three redshifts ($z=2.4$, $3.1$, and $4.5$) using three custom narrow-band (NB) filters ($N419$, $N501$, and $N673$). ODIN performs deep NB imaging with $5\sigma$ limiting magnitudes of $25.5-25.9$ over seven different wide fields, totaling a survey area of $\sim91\;\rm{deg}^2$. By combining this NB imaging with archival broad-band photometric data, ODIN will identify $>100,000$ LAEs in total. With these LAEs, we plan to study large-scale structures including protoclusters at high redshifts and their impact on galaxy evolution \citep[e.g.,][]{Ramakrishnan_2024_ApJ_filaments}. A complete description of the survey design is provided in \cite{KSLee_2024_ApJ}, and the detailed LAE selection methodology is described in \cite{Firestone_2024_ApJ_LAEselection}.

In this paper, we examine the rest-frame optical ($\sim8000\;\mathrm{\AA}$) morphologies of ODIN LAEs at redshifts $2.4$, $3.1$, and $4.5$ in the COSMOS field \citep{Scoville_2007}, using the \textit{JWST} NIRCam images from the COSMOS-Web survey \citep{Casey_2023_ApJ}. We also construct samples of typical star-forming galaxies (SFGs) at similar redshifts from the COSMOS2025 catalog \citep{Shuntov_2025_arXiv} for a robust and direct comparison between LAEs and typical SFGs. We also investigate the star formation activities of the samples using the SED fitting measurements with non-parametric SFH models from COSMOS2025 catalog. This approach allows us to systematically measure the rest-frame optical morphologies and physical properties of both ODIN LAEs and typical SFGs from COSMOS2025, enabling a robust comparison between the two populations. We note that non-parametric SFH reconstruction is particularly relevant for our analysis, as recent ODIN results with such SFH reconstruction show that most of the ODIN LAEs are undergoing their primary starburst \citep{Firestone_2025_ApJ_LAE_SFHs}.

We describe our samples of LAEs and the comparison galaxies from the COSMOS2025 catalog in Section \ref{sec:samples}. In Section \ref{sec:size_measurements}, we describe the \textit{JWST} images that we use for morphological analysis and our methodology for size measurements, and present our results and discussions in Sections \ref{sec:Results} and \ref{sec:Discussions}, respectively. Finally, we conclude and summarize our results in Section \ref{sec:Conclusions}. Throughout this study, we adopt flat $\Lambda$ Cold Dark Matter cosmological parameters of $\Omega_{m}=0.3$, $\Omega_{\Lambda}=0.7$, $\Omega_{b}=0.047$, $\sigma_8=0.816$, and $h=0.684$ as compatible with Planck data \citep{Planck_2016}.

\section{Samples} \label{sec:samples}

\subsection{ODIN LAEs} \label{subsec:ODIN_LAEs}

We utilize LAEs at redshifts $2.4$, $3.1$, and $4.5$ identified in the $\sim9 \; \rm{deg}^2$ region of the extended COSMOS field by the ODIN survey \citep{KSLee_2024_ApJ, Firestone_2024_ApJ_LAEselection}. The $5\sigma$ limiting magnitudes are $25.5$, $25.7$, and $25.9$ for the $N419$, $N501$, and $N673$ filters, respectively. The pixel size and the typical seeing of the ODIN NB images are $0\farcs27 \;{\rm{pix}^{-1}}$ and $\sim1 \arcsec$, respectively. From these NB observations, combined with archival broad-band photometry, we identify LAEs as objects with significant NB excess, which corresponds to a \Lya rest-frame equivalent width (REW) greater than $20\;\mathrm{\AA}$. We also use the NB excesses and broad-band colors to remove possible contaminators such as foreground [\ion{O}{2}] and [\ion{O}{3}] emitters. Follow-up spectroscopic observations conducted with the DESI for the $z=2.4$ and $3.1$ samples indicate that the contamination fraction for the ODIN $N419$ and $N501$ LAEs is $\lesssim10\%$ \citep{White_2024_JCAP}. A more detailed explanation and validation of the LAE selection methods can be found in \cite{Firestone_2024_ApJ_LAEselection}. 

The total number of LAE candidates identified in the extended COSMOS field is $6100$, $5782$, and $4101$ for $z=2.4$, $3.1$, and $4.5$, respectively. From this parent sample, we focus only on those within the survey area of the COSMOS-Web survey data release 1 \citep[DR1;][]{Franco_2025_arXiv}, which is used for our morphological analysis in Section \ref{sec:size_measurements}. The number of LAEs in this area is $338$, $316$, and $420$ for redshifts $2.4$, $3.1$, and $4.5$, respectively.

\subsection{Star-forming Galaxies from COSMOS2025 Catalog} \label{subsec:COSMOS2025_gals}

To compare the physical properties of our LAEs with those of other galaxy populations, we construct comparison samples of typical SFGs at redshifts similar to the ODIN LAEs. We start from the COSMOS2025 catalog \citep{Shuntov_2025_arXiv}, which covers the objects in the COSMOS-Web DR1 survey area. This catalog provides measurements of photometric redshift and various physical parameters derived from SED fitting using the \texttt{LePHARE} code \citep{Arnouts_2002_MNRAS, Ilbert_2006_A&A}, based on photometry in 37 bands with a wavelength range of $0.3-8\;\mu\mathrm{m}$. We first select the sources with photometric redshifts centered on each ODIN target redshift: i.e., $|\Delta{z_{\rm{phot}}}| < 0.2$. The \texttt{LePHARE} code also classifies each source based on its best-fit SED template type. We consider only the sources classified as ``galaxy", meaning that their best-fit template is a galaxy template. To consider only the sources with reliable photometry, we also require that there are no known problems in their photometry (i.e., \texttt{warn\_flag} = 0).

In addition to the \texttt{LePHARE} results, the COSMOS2025 catalog also provides measurements of various physical parameters, including stellar mass and SFR derived from the \texttt{CIGALE} SED fitting code \citep{Boquien_2019_A&A}. This code utilizes non-parametric SFH models, with redshifts fixed to the \texttt{LePHARE} results. Specifically, the SFH is reconstructed using ten logarithmically spaced time bins, with the first bin fixed at $10\;\mathrm{Myr}$ \citep[see][for detailed settings]{Arango-Toro_2025}. As noted by several studies \citep[e.g., ][]{Pacifici_2015_MNRAS, Iyer_2017, Carnall_2019_ApJ, Iyer_2019, Leja_2019_ApJ}, non-parametric models generally provide a more reliable reconstruction of SFHs than parametric models. This is because parametric models can impose priors on SFH reconstruction, which may bias the stellar mass or SFR measurements. Therefore, in this paper, we use the SED fitting results from \texttt{CIGALE} rather than those by \texttt{LePHARE}, which employ parametric SFH models. We refer the reader to \citet{Arango-Toro_2025} for a detailed validation of these physical parameters (stellar mass, SFR, and SFH) derived from \texttt{CIGALE}.

\begin{deluxetable*}{ll|ccccc}[t!]
\tabletypesize{\scriptsize}
    \renewcommand{\tabcolsep}{2.0mm}
    \tablecaption{Sample Sizes after Each Selection Criterion}
    \label{table:sample_sizes}
    \tablehead{
        \multicolumn{2}{l|}{\multirow{2}{*}{Sample (redshift)}} & 
        \colhead{\multirow{2}{*}{within COSMOS-Web}} & 
        \colhead{\multirow{2}{*}{$|\Delta z_{\rm{phot}}|<0.2$}} & 
        \colhead{\multirow{2}{*}{$N_{\mathrm{src}}(<0\farcs7)=1$}} & 
        \colhead{$m_{\rm{Petro}}<27.5$ and} & 
        \colhead{\multirow{2}{*}{Good \texttt{GALFIT}}} \\[-2.0ex]
        \multicolumn{2}{l|}{} & 
        \colhead{} & 
        \colhead{} & 
        \colhead{} & 
        \colhead{Point-source Rejection} & 
        \colhead{}
    }
    \renewcommand{\arraystretch}{1.5}
    \startdata
     & $(2.4)$ & $338$ & $181$ & $111$ & $92$ & $84$ \\
    LAEs & $(3.1)$ & $316$ & $189$ & $109$ & $99$ & $88$ \\
     & $(4.5)$& $420$ & $233$ & $134$ & $120$ & $75$ \\
    \hline
    & $(2.4)$ & - & $36,621$ & $23,342$ & $18,145$ & $17,949$ \\
    typical SFGs\tablenotemark{a} & $(3.1)$ & - & $20,885$ & $12,819$ & $10,273$ & $9,980$ \\
    & $(4.5)$ & - & $17,536$ & $10,160$ & $6,716$ & $3,934$ \\
    \hline
    \enddata
    \tablenotetext{a}{We do not present the number of typical SFGs within the COSMOS-Web survey area, because we started from the parent sample of SFGs as those within the area and satisfying $|\Delta z_{\rm{phot}}|<0.2$.}
\end{deluxetable*}

We select typical SFGs using a simple cut in the specific star-formation rate (sSFR) of $\log(\rm{sSFR}/\rm{yr^{-1}}) > -9.5$. This threshold is slightly lower than the sSFR of typical Lyman-break galaxies at $z>2$ \citep[see e.g., ][]{Reddy_2012, Jose_2014}. We note that using other selection criteria, such as rest-frame ${NUV-r-J}$ color-color selection commonly used in literature \citep[e.g.,][]{Ilbert_2013_A&A, Ito_2024_ApJ, Yang_2025_arXiv}, does not change the main conclusions of this paper. The number of selected typical SFGs is $36621$, $20885$, and $17536$ for $z=2.4$, $3.1$, and $4.5$, respectively.

\subsection{Final Samples} \label{subsec:final_samples}

We cross-match the ODIN LAEs with the COSMOS2025 catalog using a maximum sky separation of $1''$, and retrieve the stellar mass and SFR measurements. To secure reliable SED fitting results, we consider only the LAEs with the \texttt{LePHARE} photometric redshifts consistent with the ODIN target redshifts (i.e., $|\Delta z_{\rm{phot}}|<0.2$)\footnote{One can increase the LAE sample size by performing SED fittings of ODIN LAEs with the redshifts fixed as the ODIN target redshifts. However, this approach can result in inhomogeneous stellar mass and SFR measurements between the LAEs and typical SFGs, which could compromise a robust and systematic comparison. We therefore did not pursue this direction.}. We also reject ODIN LAEs and typical SFGs with multiple \textit{JWST} NIRCam sources within $0\farcs 7$ from the source position. This allows us to avoid any systematics of blended photometry or misidentification between ground-based (i.e., ODIN NB filters and others in COSMOS2025) and space-based observations (i.e., \textit{JWST} and \textit{HST}). These processes result in final sample sizes of $111$, $109$, and $134$ ODIN LAEs, and $23342$, $12819$, and $10160$ typical SFGs for redshifts $2.4$, $3.1$, and $4.5$, respectively. The F277W and F444W magnitude distributions of these final samples are consistent with those of their parent samples, suggesting that they are representative subsets. Therefore, we assume that this selection does not introduce significant bias in our main results. A summary of the sample sizes after each selection step is provided in Table \ref{table:sample_sizes}. We note that most of the ODIN LAEs are included in the typical SFG samples, as we do not exclude LAEs from those samples. Removing these LAEs from the typical SFG samples does not alter our main conclusions as their fraction is negligible ($\sim1\%$). This fraction is smaller than that reported in the literature, which typically ranges from $5-15\%$ for $z\sim2-4$ \citep[e.g., ][]{Stark_2010_MNRAS, Cassata_2015_A&A, Kusakabe_2020_A&A}. This could be attributed to the fact that our LAE samples are originally identified through NB filters ($\Delta z \lesssim0.08$), while typical SFGs are selected solely based on the photometric redshifts ($|\Delta z_{\rm{phot}}|<0.2$).

\begin{figure*}[t!]
    \centering
    \includegraphics[width=\textwidth]{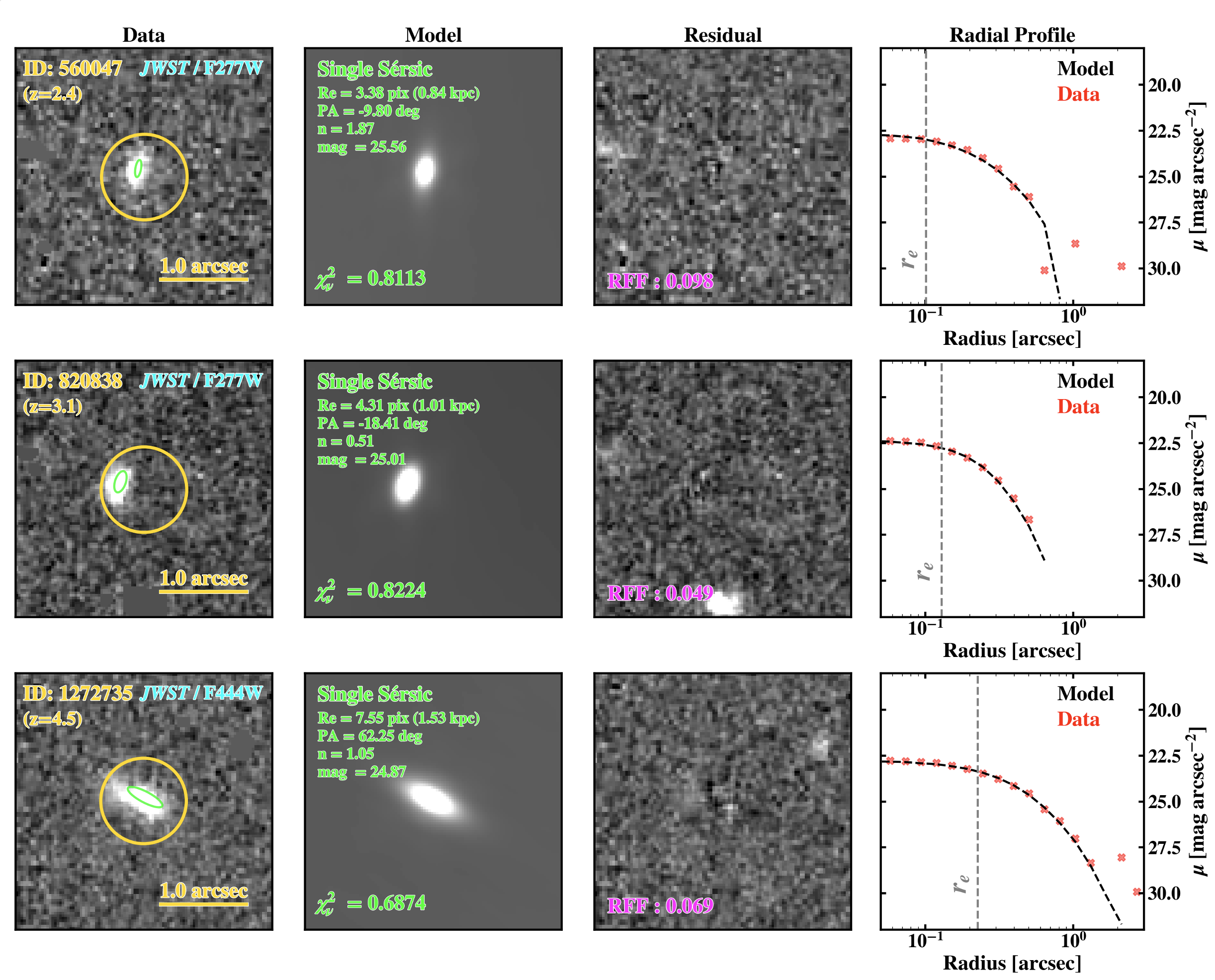}
    \caption{Representative results of the \texttt{GALFIT} fitting for ODIN LAEs at $z=2.4$, $3.1$, and $4.5$. Each row represents a case for each redshift, showing (from left to right) the masked science image (i.e., data), the modeled image with a single Sérsic profile, the residual image, and the surface brightness profile along the semi-major axis. The best-fit parameters and the residual flux fraction (RFF; see Section \ref{subsec:galfit}) are also presented. The yellow circle indicates the typical seeing size ($1''$) of ODIN NB images, while the green ellipse represents the best-fit Sérsic model.}
    \label{fig:galfit_good_results}
\end{figure*}

\section{Sizes Measurements} \label{sec:size_measurements}

\subsection{COSMOS-Web Images} \label{subsec:COSMOS-Web_images}

To study rest-frame optical $(\sim8000\rm{\AA})$ morphologies of the ODIN LAEs and the typical SFGs at $z=2.4$, $3.1$, and $4.5$, we use the \textit{JWST} NIRCam images from the COSMOS-Web DR1 \citep{Casey_2023_ApJ, Franco_2025_arXiv}. It provides mosaic images in four NIRCam filters (F115W, F150W, F277W, and F444W) with the angular pixel size of $30\;\rm{mas}$ for a contiguous $0.54\;\rm{deg}^2$ area in the COSMOS field. We choose to study rest-frame $\sim8000\rm{\AA}$ because it corresponds to the longest wavelength accessible with the F444W filter at $z=4.5$, thereby minimizing the effects of dust attenuation. We specifically utilize the images in F277W and F444W filters, with central wavelengths of $2.77$ and $4.44\;\mu m$, respectively, to study rest-frame $\sim8000 \; \rm{\AA}$ morphologies of the galaxies at $z=2.4$, $3.1$, and $4.5$.

We generate cutout images of the science and variance mosaics for the F277W and F444W filters with the sizes of $3'' \times 3''$ centered on the position of each ODIN LAE and each COSMOS2025 SFG. This size, which is $\sim20-25\;\mathrm{kpc}$ at $z=2.4-4.5$, is sufficient to contain most of the light from the galaxies in our samples. Using these science cutouts, we first run the \texttt{Source Extractor} \citep[\texttt{SExtractor;}][]{Bertin_Arnouts_1996_AnA} to identify sources in each cutout. Here, we set the \texttt{DETECT\_MINAREA} to 5 and \texttt{DETECT\_THRESH} to 2. As described in Section \ref{subsec:final_samples}, we exclude the galaxies with multiple NIRCam sources within $0\farcs7$. This allows us to securely consider the NIRCam source within $0\farcs7$ as the counterpart of each ODIN LAE and COSMOS2025 SFG, without concern about systematic effects from misidentifications or blended photometry. We then create mask images for neighboring sources (i.e., those located at $>0\farcs7$) within the cutouts using the segmentation map from \texttt{SExtractor}. To conservatively remove the effects from nearby sources, we extend the masked regions belonging to the neighbors by one pixel. Finally, we generate a bad-pixel mask for each cutout to ignore the pixels with zero variance in our morphological analyses.

To remove sources that are too faint, we exclude galaxies with a F277W Petrosian magnitude \citep{Petrosian_1976_ApJ} from \texttt{SExtractor} fainter than $27.5$. We also identify and remove unresolved point sources using the half-light radius ($r_h$) and the magnitude in F277W, based on the well-known tight correlation between the two quantities for point sources \citep[see e.g., ][]{Holwerda_2024, Ward_2024_ApJ}. We examine the relation between $r_h$ and magnitude for our sample galaxies and reject those with $r_h$ consistent with or smaller than that of point sources (see Appendix \ref{sec:point_source_rejection} for more details). The numbers of removed ODIN LAEs are $19$, $10$, and $14$ for $z=2.4$, $3.1$, and $4.5$, respectively. The corresponding numbers for the COSMOS2025 typical SFGs are $5197$, $2546$, and $3444$.

\subsection{GALFIT Runs} \label{subsec:galfit}

We utilize \texttt{GALFIT} \citep{Peng_2002_AJ, Peng_2010_AJ} to study the rest-frame optical morphologies of our sample galaxies. Here, we use the cutouts of the science and variance images, along with mask images for both nearby sources and bad pixels. We adopt a single Sérsic profile \citep{Sérsic_1963_BAAA} to model each science cutout and measure the effective radius ($R_e$) and Sérsic index, as in previous studies \citep[e.g., ][]{LeeJH_2024_ApJ, Ward_2024_ApJ, Martorano_2024_ApJ, Yang_2025_arXiv}. For the samples at $z=2.4$ and $4.5$, we use the F277W and F444W cutouts to measure rest-frame optical ($\sim8000\;\rm{\AA}$) sizes and Sérsic indices, respectively. For the samples at $z=3.1$, we independently measure these parameters using both filters and perform a simple linear interpolation to infer their values at rest-frame $\sim8000\;\rm{\AA}$. Here, we assume that the size and the Sérsic index of a galaxy change smoothly as a function of wavelength, and can be modeled by simple polynomials \citep[see e.g., ][]{Ward_2024_ApJ, Martorano_2024_ApJ}.

When \texttt{GALFIT} finds the best-fit model, it compares the observed image to the modeled image convolved with a point-spread function (PSF). Recent studies of high-redshift galaxy morphologies based on \textit{JWST} NIRCam mosaic images \citep[e.g., ][]{LeeJH_2024_ApJ, Martorano_2024_ApJ} report that the PSFs directly obtained from the mosaics are slightly broader than those modeled with the \texttt{WEBBPSF} \citep{Perrin_2014_SPIE}. Additionally, some studies report that PSFs can have spatial variations as well \citep[e.g., ][]{Zhuang_2024_ApJ_962}. To avoid any systematic biases of using an inappropriate PSF, we empirically select stars and construct our own PSFs using the \texttt{PSFEx} \citep{Bertin_2011_ASPC} software, as done in other studies \citep[e.g., ][]{Zhuang_2024_ApJ_962, LeeJH_2024_ApJ}. Because the entire COSMOS-Web DR1 data are divided into 20 different mosaics, we select stars and generate a PSF model for each filter and mosaic combination. The number of selected stars in each mosaic is $\sim170-300$ for F277W and $\sim80-140$ for F444W filter. Finally, we use the PSF model generated from each mosaic for the \texttt{GALFIT} fitting of the sample galaxies within the same mosaic. 

During the fitting process, we allow the source center to move by $\pm3$ pixels from the source position determined by the \texttt{SExtractor}. We impose the constraints on the effective radius of $[0.1,50.]$ pixels, and on the Sérsic index of $[0.1,10.]$. We also impose a constraint on the total magnitude of the best-fit model to be within $\pm0.5\;\rm{mag}$ from the Petrosian magnitude \citep{Petrosian_1976_ApJ} obtained with \texttt{SExtractor}.

To select only the sample galaxies with reliable \texttt{GALFIT} results, we use the residual flux fraction \citep[RFF;][]{Hoyos_2011_MNRAS, Hoyos_2012_MNRAS}, which measures the amount of the signal in the residual image that cannot be explained by background fluctuation. As in \cite{LeeJH_2024_ApJ}, we calculate RFF using the following equation:
\begin{equation}
    \mathrm{RFF}=\frac {\Sigma_{i,j}|I(i,j)^{\mathrm{residual}}| - 0.8\times\Sigma_{i,j}\sigma_B(i,j)} {\Sigma_{i,j} I(i,j)^{\mathrm{data}}}.
\end{equation}
Here, $I(i,j)^\mathrm{data}$ and $I(i,j)^\mathrm{residual}$ are the pixel values in the original and the residual images, and $\sigma_{B}(i,j)$ represents the background fluctuation in each pixel. The summation is performed over a certain area around each source which we define as an elliptical region with a semimajor axis of $1.5 \times r_P$, where $r_P$ is the Petrosian radius from the \texttt{SExtractor}. The second term in the numerator, with a factor of $0.8$, represents the expected amount of the signal when assuming that the sky background in each pixel follows a Gaussian distribution with the mean of zero and the standard deviation of $\sigma_B$. We estimate $\sigma_B$, by calculating the standard deviation with sigma clipping of the pixel values outside the  $1.5 \times r_P$ region of all the sources in each cutout. We then impose a maximum RFF of $0.5$ as in previous studies \citep[e.g., ][]{Ormerod_2024_MNRAS, Ward_2024_ApJ, LeeJH_2024_ApJ}.

We exclude galaxies whose $R_e$ or Sérsic index did not converge well. We further reject those with the error of the Sérsic index is greater than the best-fit value (i.e., $\delta n > n$). However, we do not exclude the galaxies with $R_e$ smaller than the PSF size (i.e., $R_e < 0.5 \times \mathrm{FWHM}$), while their measured sizes could be slightly overestimated by $\lesssim0.1\;\mathrm{dex}$ \citep[see e.g., ][]{Davari_2014, Sun_2024, Ito_2024_ApJ}. This would not alter our main results of comparative analysis between ODIN LAEs and typical SFGs in Sections \ref{subsec:rest-optical_morphologies} and \ref{subsec:SFMS}.

The numbers of ODIN LAEs with reliable \texttt{GALFIT} measurements in the final sample are $84$, $88$, and $75$ at $z=2.4$, $3.1$, and $4.5$, respectively; the corresponding numbers of typical SFGs are $17949$, $9980$, and $3934$. We examine the F277W magnitude distributions and find that this selection preferentially leaves brighter galaxies, effectively acting as a brighter survey limit for the NIRCam images. We do not expect this to introduce significant bias in our results, particularly in the comparative analysis in Sections \ref{subsec:rest-optical_morphologies} and \ref{subsec:SFMS}, where we compare samples of LAEs and SFGs that are subject to similar selection effects. Figure \ref{fig:galfit_good_results} shows three representative examples of the \texttt{GALFIT} results for ODIN LAEs at $z=2.4$, $3.1$, and $4.5$. The resulting morphological parameters, along with the physical properties from the ODIN and COSMOS2025 catalogs, for ODIN LAEs are available in Table \ref{table:GALFIT_results}.

\setlength{\arrayrulewidth}{0.6pt}
\begin{deluxetable*}{lccccccccc}[t!]
    \tabletypesize{\scriptsize}
    \renewcommand{\arraystretch}{1.7}
    \renewcommand{\tabcolsep}{2.0mm}
    \tablecaption{GALFIT results of ODIN LAEs}
    \label{table:GALFIT_results}
    \digitalasset
    \tablehead{\colhead{ID} & \colhead{ODIN filter} & \colhead{R.A.} & \colhead{Decl.} & \colhead{$R_e$} & \colhead{$n_{\text{Sérsic}}$} & \colhead{\Lya REW} & \colhead{$M_{\star}$} & \colhead{$\text{SFR}_{\text{inst}}$} & \colhead{$\text{SFR}_{\text{100Myr}}$} \\
    \colhead{} & \colhead{} & \colhead{[deg]} & \colhead{[deg]} & \colhead{[kpc]} & \colhead{} & \colhead{[$\mathrm{\AA}$]} & \colhead{[$10^9 \; M_{\odot}$]} & \colhead{[$M_{\odot}\;\text{yr}^{-1}$]} & \colhead{[$M_{\odot}\;\text{yr}^{-1}$]} \\ 
    \colhead{(1)} & \colhead{(2)} & \colhead{(3)} & \colhead{(4)} & \colhead{(5)} & \colhead{(6)} & \colhead{(7)} & \colhead{(8)} & \colhead{(9)} & \colhead{(10)}}
    \startdata
    $530967$ & $N419$ & $150.33026$ & $1.79015$ & $0.509\pm0.008$ & $0.793\pm0.089$ & $29.334\pm6.997$ & $0.987\pm0.066$ & $0.938\pm0.047$ & $3.798\pm0.512$ \\
    $535278$ & $N419$ & $150.26980$ & $1.83013$ & $0.963\pm0.009$ & $0.656\pm0.035$ & $40.771\pm5.918$ & $1.140\pm0.081$ & $1.872\pm0.094$ & $7.552\pm0.378$ \\
    $537715$ & $N419$ & $150.12939$ & $1.85398$ & $0.953\pm0.022$ & $0.449\pm0.059$ & $30.292\pm8.722$ & $0.976\pm0.074$ & $1.340\pm0.293$ & $3.419\pm0.335$ \\
    \multicolumn{10}{c}{$\vdots$} \\
    $728259$ & $N501$ & $150.28549$ & $1.73716$ & $1.666\pm2.151$ & $4.710\pm3.181$ & $95.469\pm12.891$ & $0.257\pm0.054$ & $2.233\pm0.348$ & $1.338\pm0.347$ \\
    $738334$ & $N501$ & $150.10568$ & $1.80673$ & $0.322\pm0.076$ & $4.476\pm6.526$ & $77.116\pm11.344$ & $0.259\pm0.042$ & $2.507\pm0.252$ & $1.518\pm0.286$ \\
    $740729$ & $N501$ & $150.22230$ & $1.82227$ & $1.238\pm0.069$ & $3.359\pm0.254$ & $63.026\pm7.294$ & $5.249\pm0.409$ & $4.075\pm1.089$ & $12.415\pm3.265$ \\
    \multicolumn{10}{c}{$\vdots$} \\
    $1250880$ & $N673$ & $150.19823$ & $1.82594$ & $2.074\pm1.782$ & $5.312\pm4.999$ & $68.624\pm19.165$ & $0.845\pm0.505$ & $5.255\pm1.278$ & $3.099\pm1.511$ \\
    $1267104$ & $N673$ & $150.20686$ & $1.89213$ & $0.257\pm0.068$ & $4.553\pm4.324$ & $144.011\pm19.833$ & $0.590\pm0.037$ & $2.041\pm0.411$ & $4.581\pm1.438$ \\
    $1267583$ & $N673$ & $150.31278$ & $1.89406$ & $1.346\pm0.054$ & $1.463\pm0.113$ & $29.935\pm9.146$ & $2.803\pm0.140$ & $3.339\pm0.167$ & $23.956\pm1.198$ \\
    \multicolumn{10}{c}{$\vdots$} \\
    \hline
    \enddata
    \tablecomments{(1) ODIN LAE ID, (2) ODIN NB filter where each LAE is identified, (3)-(4) coordinates on the ODIN NB image, (5) Effective radius measured with \texttt{GALFIT}, (6) Sérsic index measured with \texttt{GALFIT}, (7) rest-frame equivalent width of the \Lya emission line, (8)-(10) stellar mass, instantaneous SFR, and SFR averaged over the last $100$ Myr from the COSMOS2025 catalog.}
    \tablecomments{We present only a portion of the entire table, which is available in machine-readable format.}
\end{deluxetable*}

\section{Results} \label{sec:Results}

\subsection{Rest-frame Optical Morphologies} \label{subsec:rest-optical_morphologies}

\subsubsection{Size-mass Relations} \label{subsec:size-mass_relations}

\begin{figure*}[t!]
    \centering
    \includegraphics[width=\textwidth]{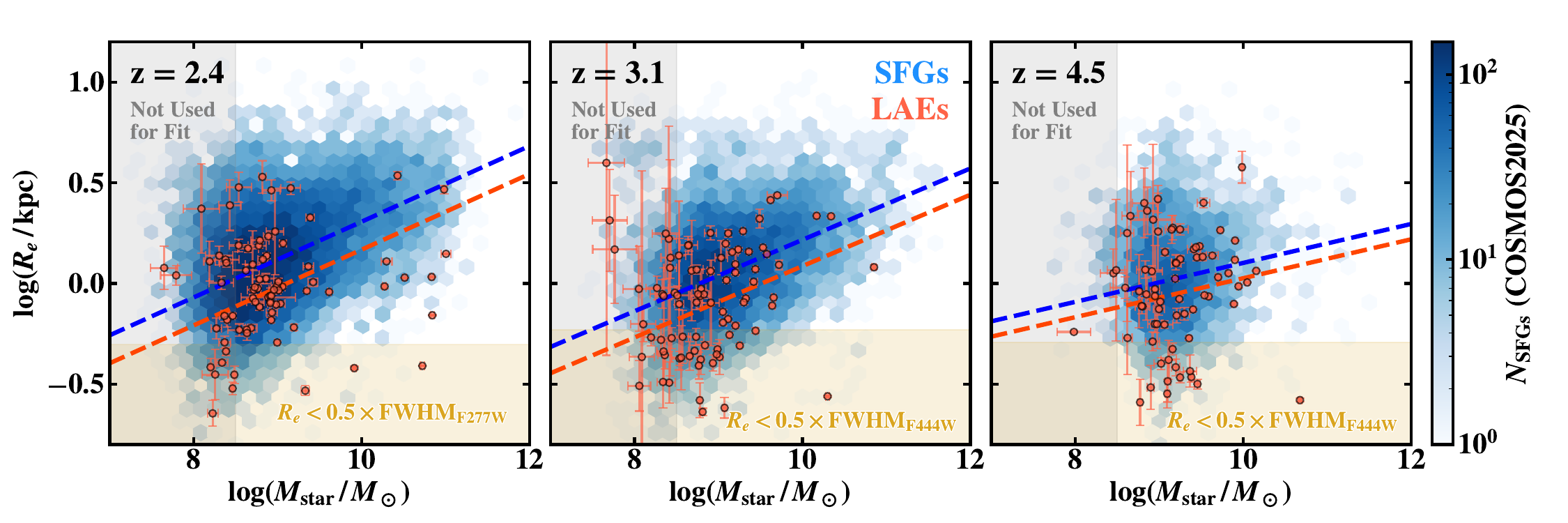}
    \caption{Rest-frame optical ($\sim8000\;\mathrm{\AA}$) size-mass relations of LAEs and typical SFGs at $z=2.4$, $3.1$, and $4.5$. LAEs are shown as data points with error bars, while typical SFGs are represented by the background color map. In each panel, the best-fit power-law relations for LAEs and typical SFGs are presented as the red and blue dashed lines, respectively. Some of the sample galaxies have $R_e$ smaller than the PSF size (i.e., $R_e<0.5 \times$ FWHM), while we do not exclude them from our analysis (see Section \ref{subsec:galfit}).}
    \label{fig:size-mass}
\end{figure*}

We present the size-mass relations of the ODIN LAEs and the typical SFGs in Figure \ref{fig:size-mass}, along with the best-fit power-law relation. We fit the size-mass relation of the typical SFGs using a single power-law function with an intrinsic scatter in size as commonly adopted for high-redshift galaxies \citep[e.g., ][]{vdWel_2014_ApJ, Ito_2024_ApJ, Nedkova_2024_ApJ, Yang_2025_arXiv}. In this model, the sizes of galaxies at a fixed stellar mass follow a log-normal distribution with $\log{R_e} \sim N(\langle \log{R_e} \rangle, \sigma_{\rm{int}})$, and the mean is related to stellar mass by a single power-law as follows:
\begin{equation}
    \frac{\langle R_e \rangle}{\rm{kpc}} = A { \left( \frac{M_{\star}}{5\times10^{10}\;M_{\odot}} \right)} ^{\beta}.
    \label{eq:size-mass-formula}
\end{equation}
Here, $A$ and $\beta$ represent the mean size of the galaxies at $M_{\star}=5\times10^{10}\;M_{\odot}$ and the power-law slope, respectively. To find the best-fit relations, we use the Markov Chain Monte Carlo (MCMC) method using the \texttt{emcee} Python package \citep{Foreman-Mackey_2013_PASP}. In the fitting process, we only use the SFGs with $\log(M_{\star}/M_{\odot}) > 8.5$ to consider only the galaxies above the stellar mass completeness limit of the COSMOS2025 catalog (see Figure~24 of \citealt{Shuntov_2025_arXiv}). We note that imposing an upper limit on ${M_{\star}}$ for the fitting range\footnote{We perform this test due to concerns that the small number of galaxies at the high-mass end could potentially alter the slope of the size-mass relation.} or utilizing size-mass relations of SFGs from the literature \citep[e.g., ][]{Yang_2025_arXiv} does not change the main conclusions of this paper. However, we do not apply this stellar mass cut in the subsequent analysis, assuming that typical SFGs and LAEs below the completeness limit are subject to similar selection effects.

\begin{figure*}[t!]
    \centering
    \includegraphics[width=\textwidth]{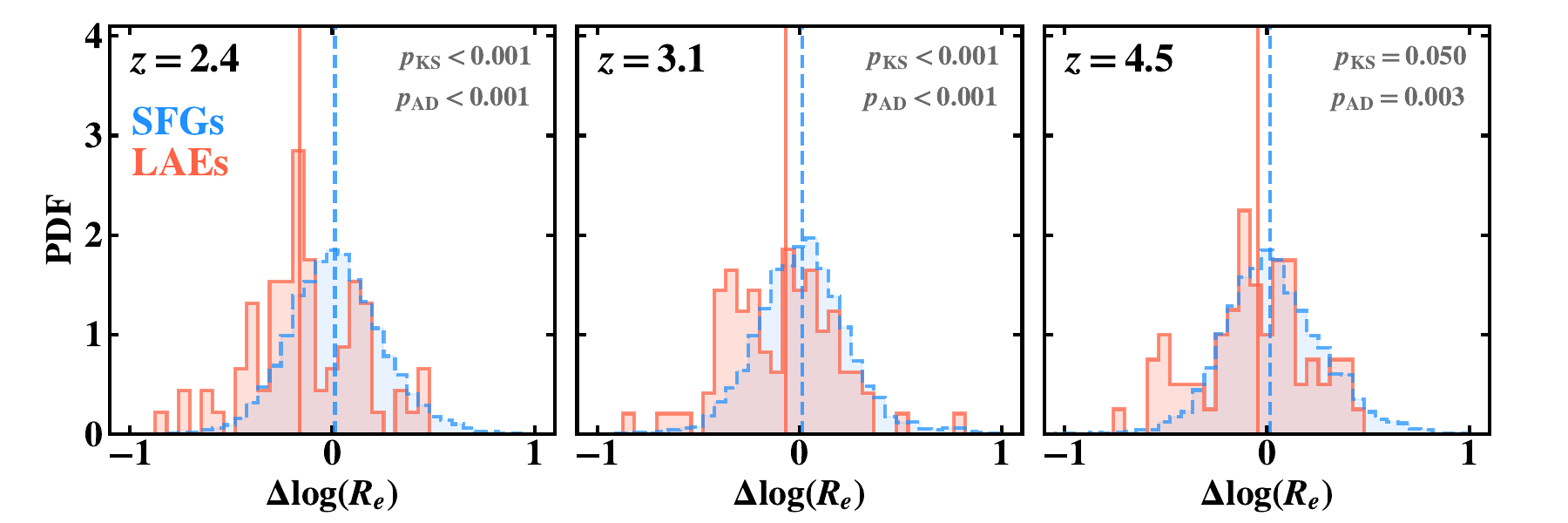}
    \caption{Distribution of the size difference ($\Delta \log R_e$) between each galaxy and the best-fit size-mass relations. The median values for LAEs and typical SFGs at each redshift are shown as solid and dashed lines, respectively. We also present the $p$-values from the KS- and AD tests for the distributions at each redshift. LAEs tend to have smaller sizes than typical SFGs at $z=2.4$ and $3.1$, while there is weaker difference at $z=4.5$.}
    \label{fig:logD_hists}
\end{figure*}

We construct the likelihood function by considering uncertainties in both stellar mass and effective radius. Following the approach of \cite{vdWel_2014_ApJ} and \cite{Yang_2025_arXiv}, we convert the uncertainties in stellar mass measurements into those in size, by assuming a typical slope of the size-mass relations of $0.2$. We set the initial guesses for the parameters $A$ and $\beta$ as the best-fit values obtained with simple least-squares fitting, while the intrinsic scatter $\sigma_{\rm{int}}$ is initialized to $0.2$. We assume flat priors for these parameters within the ranges $[0,10]$, $[0,5]$, and $[0,2]$ for $A$, $\beta$, and $\sigma_{\rm{int}}$, respectively. For the MCMC analysis, we use $100$ walkers with $2,500$ steps, discarding the first $500$ steps. The resulting best-fit power-law relation for typical SFGs at each redshift is represented as the blue dashed line in each panel of Figure \ref{fig:size-mass}. The best-fit parameters are available in Table \ref{table:best-fit_valse}.

\begin{figure*}[t!]
    \centering
    \includegraphics[width=\textwidth]{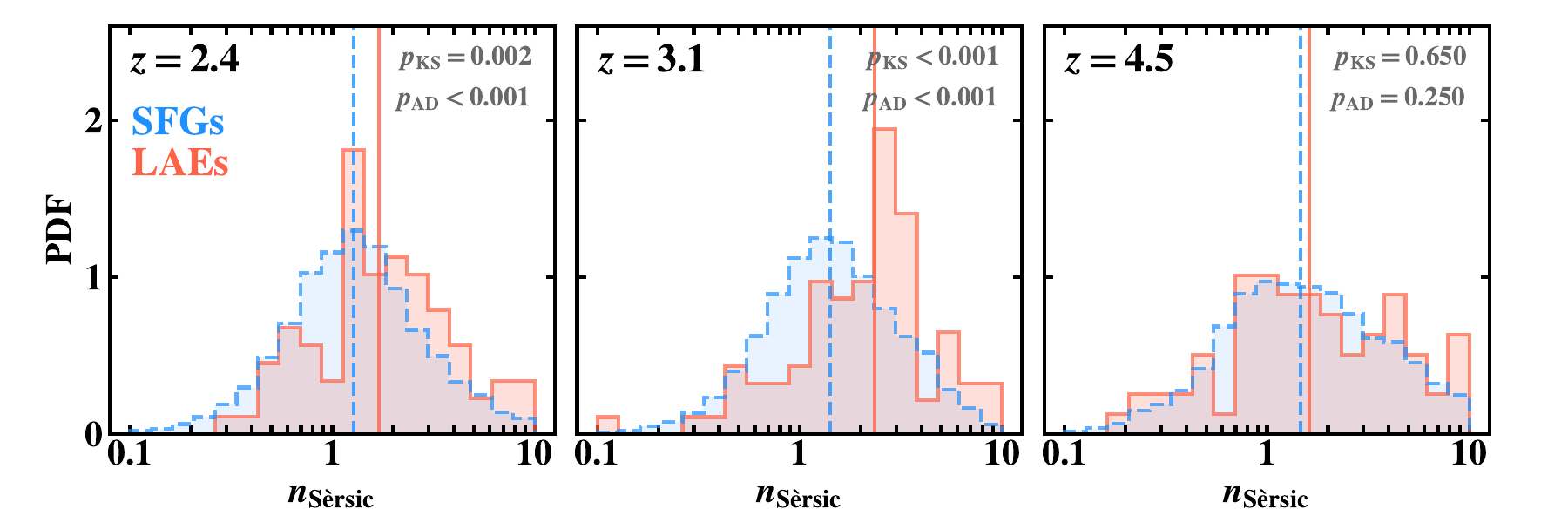}
    \caption{Distributions of Sérsic indices of ODIN LAEs and the typical SFGs from COSMOS2025 catalog. The median Sérsic indices of LAEs and typical SFGs at each redshift are represented as solid and dashed lines, respectively. The $p$-values from the KS- and AD tests for each redshift are available in the upper right corner of each panel. LAEs at $z=2.4$ and $3.1$ shows larger Sérsic index than typical SFGs at similar redshifts, while the difference becomes smaller and weaker at $z=4.5$.}
    \label{fig:nSersic_hists_ODIN_LAEs}
\end{figure*}

To compare the sizes of LAEs and typical SFGs at similar redshifts without the effect of stellar mass, we measure the size excess ($\Delta \log R_e$) from the best-fit size-mass relation using the following equation:
\begin{equation}
    \Delta \log R_e=\log \left( \frac{R_e}{\langle R_e \rangle(M_{\star})} \right),
    \label{eq:logD_formula}
\end{equation}
where $\langle R_e \rangle (M_{\star})$ is the mean effective radius at a given stellar mass. We present the histogram of $\Delta \log R_e$ in Figure \ref{fig:logD_hists} along with the median for each sample\footnote{Although the size distributions of LAEs show hints of bimodality, our statistical analysis using Gaussian Mixture Models and the Bayesian Information Criterion favors a single component model for all three redshifts.}. LAEs at $z=2.4$ and $3.1$ tend to have smaller sizes than typical SFGs at similar redshifts, but those at $z=4.5$ show less evidence for size difference. To statistically determine whether the two samples are drawn from the same underlying size distribution, we perform the Kolmogorov–Smirnov (KS) and Anderson–Darling (AD) k-sample tests on the two histograms at each redshift. The resulting $p$-values for both tests are $<0.001$ at $z=2.4$ and $3.1$, indicating that LAEs have distinct size distributions compared with typical SFGs at similar redshifts. The $p$-values for the case of $z=4.5$ are still small, but is not as significant as those for $z=2.4$ and $3.1$ (i.e., $\gtrsim2\sigma$ level). To verify that the small $p$-values are not driven by the large difference in sample sizes between LAEs and typical SFGs, we repeat the tests using random subsets of typical SFGs matched to the number of LAEs at each redshift. We confirm that this does not alter our main conclusions. This trend will be examined using the Horizon Run 5 cosmological simulation in Section \ref{subsec:size-mass_inHR5}.

\subsubsection{Sérsic Index}\label{subsec:sersic_indices}

We compare the distributions of Sérsic indices for ODIN LAEs and the typical SFGs in Figure \ref{fig:nSersic_hists_ODIN_LAEs}. We also perform KS- and AD tests for the two distributions at each redshift. We find that LAEs at $z=2.4$ and $3.1$ tend to have larger Sérsic indices than typical SFGs at similar redshifts. However, at $z=4.5$, the differences become smaller and less significant (see the $p$-values in the upper right corner of each panel). In Section \ref{subsec:size-mass_inHR5}, we examine the distributions of Sérsic indices for the LAEs and SFGs in the Horizon Run 5 cosmological simulation. Although we investigated the relationship between the stellar mass and Sérsic index, we could not find a clear correlation between the two. Therefore, we only present the distributions of the Sérsic indices themselves, without correcting their (possible) stellar-mass dependence. We note that comparing Sérsic indices of the two populations after normalizing the stellar-mass effect with the best-fit mass-Sérsic index relations (as similar to what we did for the sizes in Section \ref{subsec:size-mass_relations}) yields results consistent with those shown in Figure \ref{fig:nSersic_hists_ODIN_LAEs}.

\begin{figure*}[t!]
    \centering
    \includegraphics[width=\textwidth]{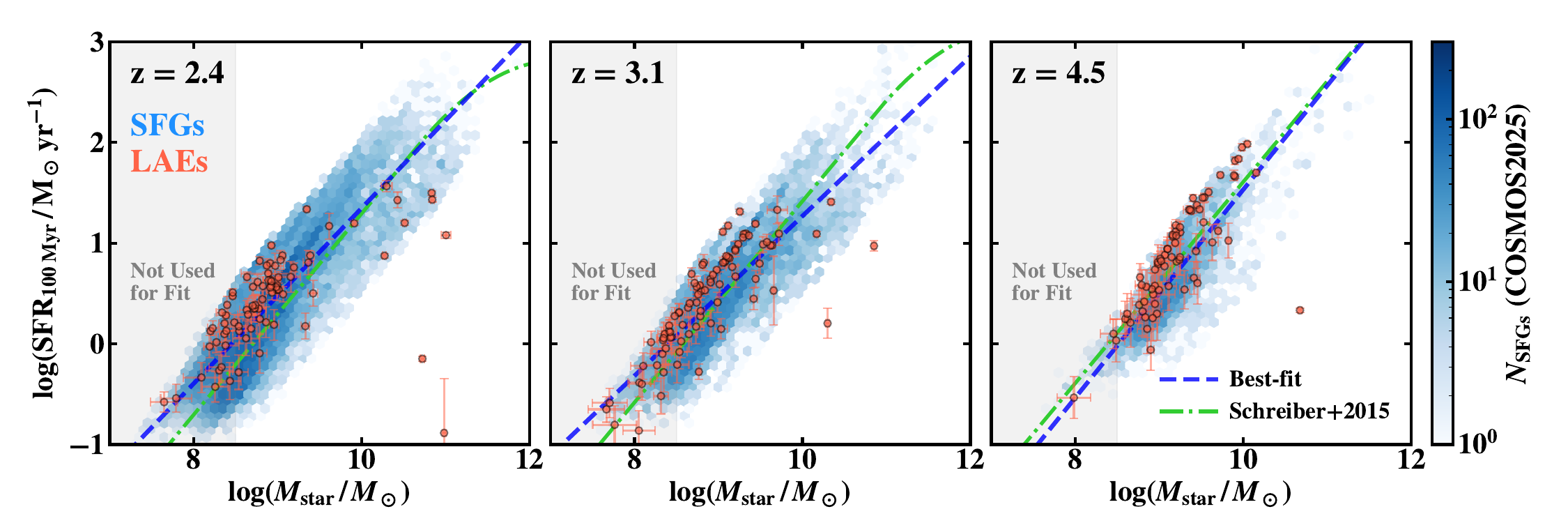}
    \caption{Relations between stellar mass and star-formation rate of ODIN LAEs and the star-forming galaxies from COSMOS2025 catalog at $z=2.4$, $3.1$, and $4.5$. The best-fit relations from our MCMC fitting along with those from \cite{Schreiber_2015_A&A} for each redshift is presented in each panel. LAEs tend to be located above this line.}
    \label{fig:SFMS}
\end{figure*}

\begin{figure*}[t!]
    \centering
    \includegraphics[width=\textwidth]{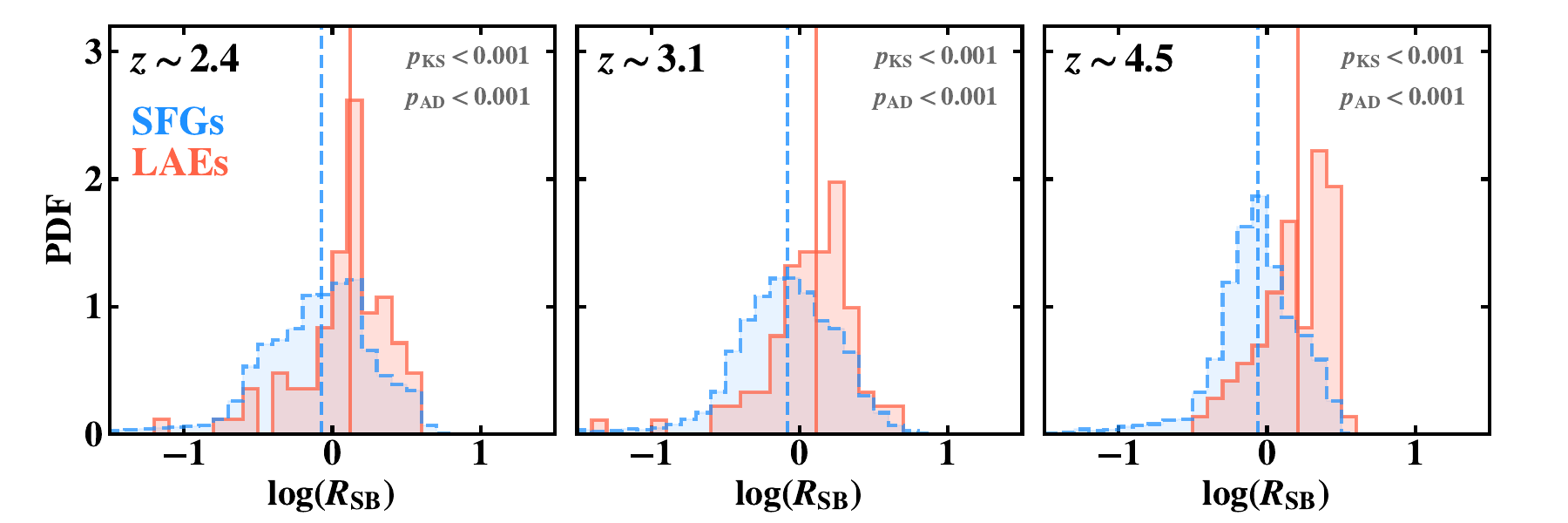}
    \caption{Distributions of starburstiness $(R_{\rm{SB}})$ for ODIN LAEs and the typical SFGs at $z=2.4$, $3.1$, and $4.5$. The $R_{\rm{SB}}$ is defined by Equation (\ref{eq:starburstiness}). LAEs have higher $R_{\rm{SB}}$ at all three redshifts, indicating they are preferentially in a starburst phase.}
    \label{fig:RSB_histogram}
\end{figure*}

\begin{figure*}[t!]
    \centering
    \includegraphics[width=\textwidth]{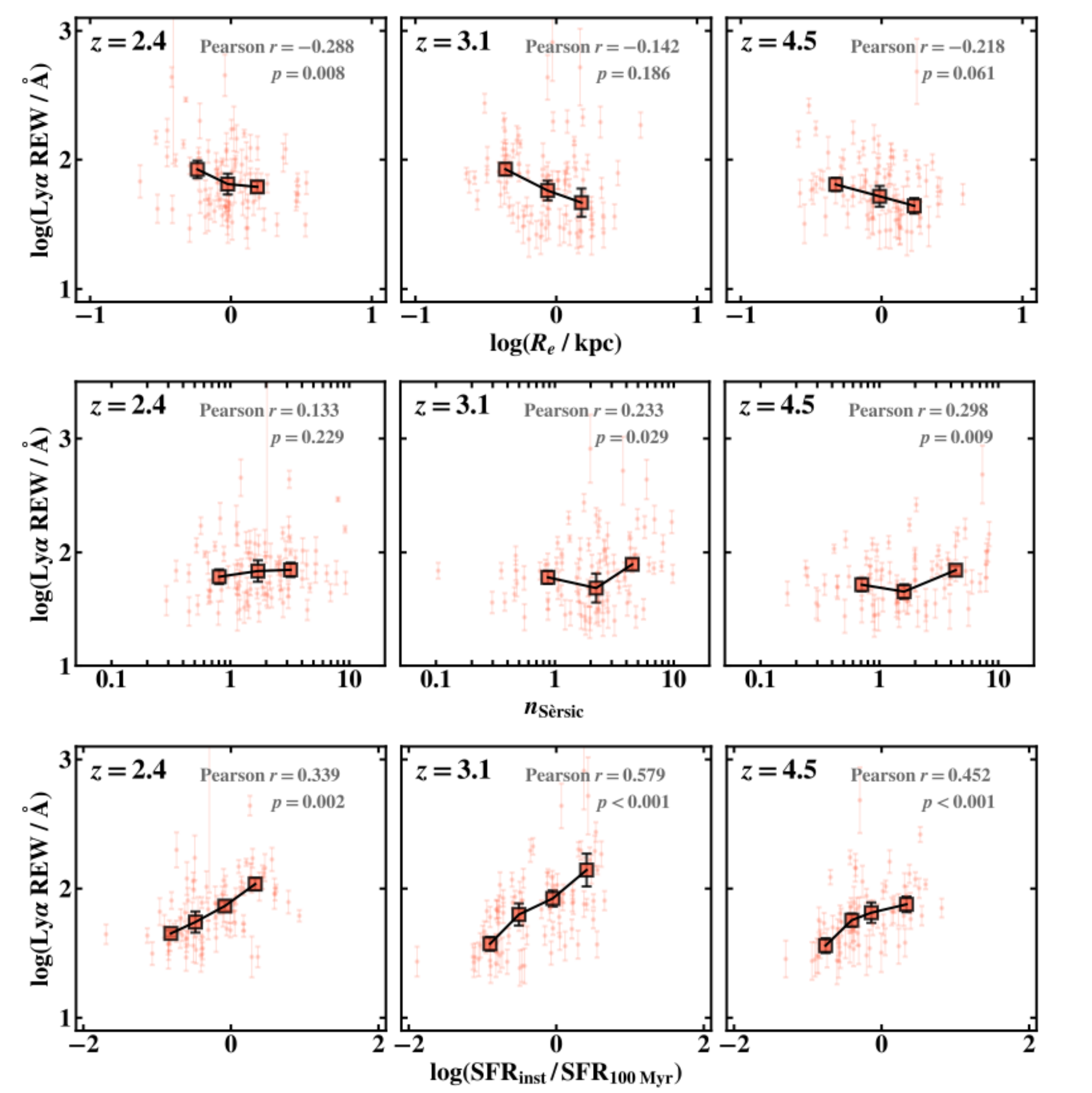}
    \caption{Correlations between the rest-frame equivalent width (REW) of \Lya emission line and galaxy properties for ODIN LAEs. From top to bottom, the panels show the relations with rest-frame optical sizes, Sérsic indices, and the ratio of the instantaneous SFR to the SFR averaged over the last $100\;\mathrm{Myr}$ (i.e., $\mathrm{SFR}_{\mathrm{inst}}/\mathrm{SFR}_{\mathrm{100\,Myr}}$). We also present the Pearson correlation coefficient and the corresponding $p$-value in each panel. Square data points represent the median REW for each bin of galaxy properties, along with their standard deviations, which are calculated by bootstrapping with $1,000$ times.}
    \label{fig:Lya_REW_correlations}
\end{figure*}

\subsection{Star-forming Main-sequences} \label{subsec:SFMS}

To compare the star formation activities of LAEs and typical SFGs, we examine the relations between stellar masses and SFRs, known as the star-forming main-sequence (SFMS). For this analysis, we adopt SFR averaged over the last $100\;\mathrm{Myr}$, derived from the \texttt{CIGALE} SED fitting of the COSMOS2025 catalog. We assume that the SFMS follows a power-law relation with intrinsic scatter, as in previous studies \citep[e.g., ][]{Kurczynski_2016, Simmonds_2025}. We perform MCMC fitting as similar to that in Section \ref{subsec:size-mass_relations}, while we adopt a typical SFMS slope of 1.0 to convert the stellar-mass uncertainties into those in SFR. Figure \ref{fig:SFMS} shows the resulting best-fit relations, along with the SFMS from the literature \citep[][]{Schreiber_2015_A&A}. For the subsequent analysis, we utilize the SFMS relations derived from our MCMC fitting; however, we note that using the relations of \cite{Schreiber_2015_A&A} does not alter our main results. LAEs are consistently located above the main-sequence line at all three redshifts. To compare the SFRs by removing the stellar mass effects, we calculate the starburstiness \citep[$R_{\rm{SB}}$;][]{Elbaz_2011_A&A} of each LAE and SFG using the following equation:
\begin{equation}\label{eq:starburstiness}
    R_{\rm{SB}} = \frac{\rm{SFR}}{\rm{SFR}_{MS}},
\end{equation}
where $\rm{SFR}$ is the star-formation rate of each sample galaxy and ${\rm{SFR}}_{\rm{MS}}$ is that of the best-fit SFMS relation. We present the distributions of the $R_{\rm{SB}}$ for our samples in Figure \ref{fig:RSB_histogram} along with the median for each sample. We also perform the KS- and AD tests on the two distributions at each redshift and present the resulting $p$-values as well. At all three redshifts, we find that LAEs tend to have higher $R_{\rm{SB}}$ than typical SFGs at similar redshifts, with high statistical significance. This indicates that LAEs are generally more active in star formation than typical SFGs; this is consistent with recent results from observations \citep[e.g., ][]{Liu_2024_ApJ, Ning_2024_ApJ} and cosmological simulations \citep[e.g., ][]{Im_2024_ApJ}. This also aligns with recent findings from the ODIN survey, which suggest that the ODIN LAEs are undergoing the most significant starburst in their SFHs \citep{Firestone_2025_ApJ_LAE_SFHs}.

Although our LAEs exhibit enhanced SFRs in general, we note that our data also suggest a stellar-mass dependence. At the high-mass regime ($\log (M_{\star}/M_{\odot}) > 10$), the majority of our LAEs are located below the SFMS, which aligns with recent findings that high-mass LAEs tend to lie on or even below the main sequence \citep[e.g., ][]{Santos_2020_MNRAS, Goovaerts_2024_A&A}. Furthermore, \cite{McCarron_2022_ApJ} reported a result that low-mass ($\log (M_{\star}/M_{\odot}) < 9$) LAEs tend to be located below the SFMS. These findings may imply that the \Lya production and/or escape mechanisms could vary across different stellar-mass regimes. To fully understand this stellar-mass dependence, a more detailed analysis of the physical properties of these LAEs is necessary.

\subsection{Correlation between \Lya REW and galaxy properties} \label{subsec:REW_correlations}

We examine the relation between galaxy properties and the rest-frame equivalent width of \Lya emission line (REW) for ODIN LAEs. The \Lya REW is calculated using a combination of the narrow- and broad-band photometry (see Section 3 of \citealt{Firestone_2024_ApJ_LAEselection} for more details). Because the \Lya escape fraction ($f_{\mathrm{esc}}^{\mathrm{Ly}\alpha}$) is known to be positively correlated with the \Lya REW \citep[e.g., ][]{Sobral_2019_A&A, Jones_2025_MNRAS}, these correlations would provide important insights into the physical conditions related to \Lya photon escape from galaxies. 

We first examine the relation between the rest-frame optical sizes and the \Lya REW (top panels of Figure \ref{fig:Lya_REW_correlations}). To visualize this relation, we bin the data with respect to the rest-frame optical size such that each bin contains the same number of LAEs, and present the median \Lya REW for each bin as well. To have statistical interpretations, we calculate and present the Pearson correlation coefficients and the corresponding $p$-values. Although the statistical significance is relatively small at $z=3.1$, we find hints of negative correlations between \Lya REW and the rest-frame optical size at all three redshifts. This could indicate that smaller galaxies tend to have higher $f_{\mathrm{esc}}^{\mathrm{Ly}\alpha}$. The middle panels show hint of positive correlations between the Sérsic index and \Lya REW at $z=3.1$ and $4.5$, as supported by the Pearson statistics. This can suggest that galaxies with higher central concentrations tend to have higher $f_{\mathrm{esc}}^{\mathrm{Ly}\alpha}$. We discuss more about these findings in Section \ref{subsec:Lya_escape_conditions}. 

We also utilize the non-parametric reconstruction of star-formation histories for ODIN LAEs, derived from the \texttt{CIGALE} SED fittings of the COSMOS2025 catalog. Specifically, we focus on the ratio of the instantaneous SFR and that averaged over the last $100\;\mathrm{Myr}$ (i.e., $\mathrm{SFR}_{\mathrm{inst}}\,/\,\mathrm{SFR}_{\mathrm{100\;Myr}}$). This estimates how active the current star formation is, compared with the last $100\;\mathrm{Myr}$. In Figure \ref{fig:Lya_REW_correlations} (bottom panels), we present correlations between \Lya REW and the $\mathrm{SFR}_{\mathrm{inst}}\,/\,\mathrm{SFR}_{\mathrm{100\;Myr}}$, along with the Pearson correlation coefficients and corresponding $p$-values. At all three redshifts, we find strong positive correlations between the two. We note that the \Lya REW and $\mathrm{SFR}_{\mathrm{inst}}\,/\,\mathrm{SFR}_{\mathrm{100\;Myr}}$ were estimated independently of each other. This is expected because the intrinsic \Lya REW (i.e., before experiencing the resonant scattering) is directly proportional to the ratio of the SFR over the past $<10\;\mathrm{Myr}$ to that averaged over $100\;\mathrm{Myr}$ \citep[e.g., ][]{Dijkstra_2010}.

These correlations between the physical properties and \Lya REW can provide important insights into the physical conditions related to the \Lya photon escape. We give further discussion of these findings in Section \ref{subsec:Lya_escape_conditions}.

\begin{figure*}[ht!]
    \centering
    \includegraphics[width=\textwidth]{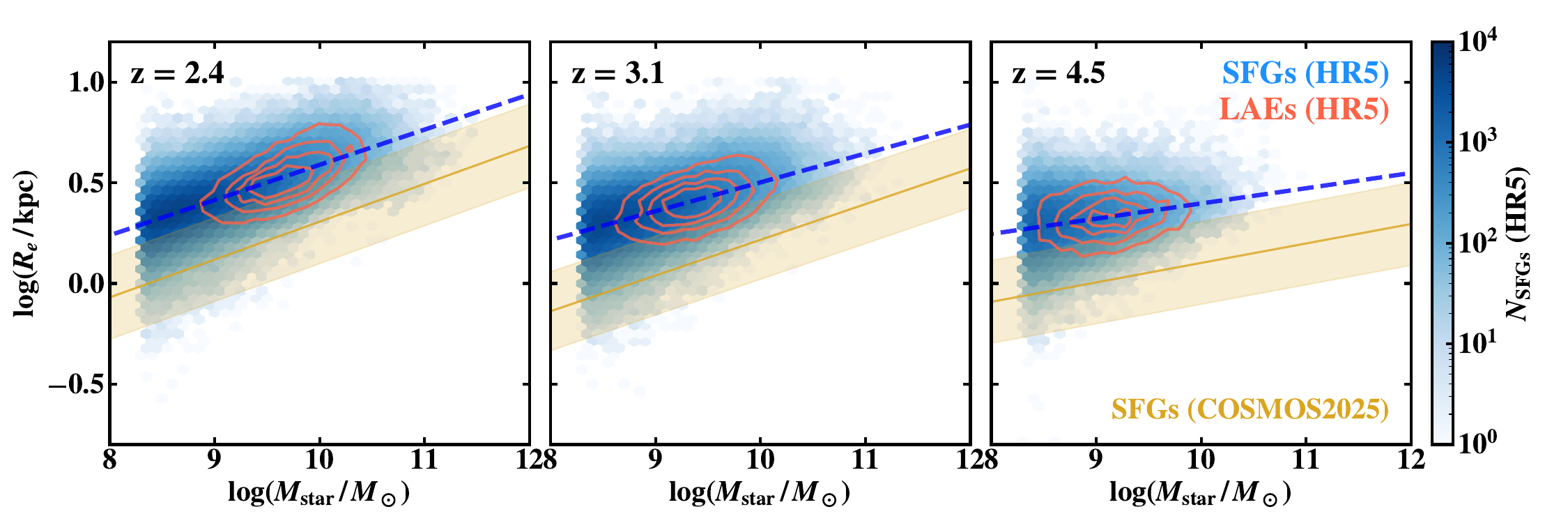}
    \caption{Size-mass relations of LAEs (red contours) and typical SFGs (background color map) in the Horizon Run 5 cosmological hydrodynamical simulation. The blue dashed line in each panel represents the best-fit power-law relation for typical SFGs at each redshift. For comparison, we also present the best-fit size-mass relations for the typical SFG samples from COSMOS2025 catalog (see Section \ref{subsec:size-mass_relations}).}
    \label{fig:size-mass_inHR5}
\end{figure*}

\begin{figure*}[ht!]
    \centering
    \includegraphics[width=\textwidth]{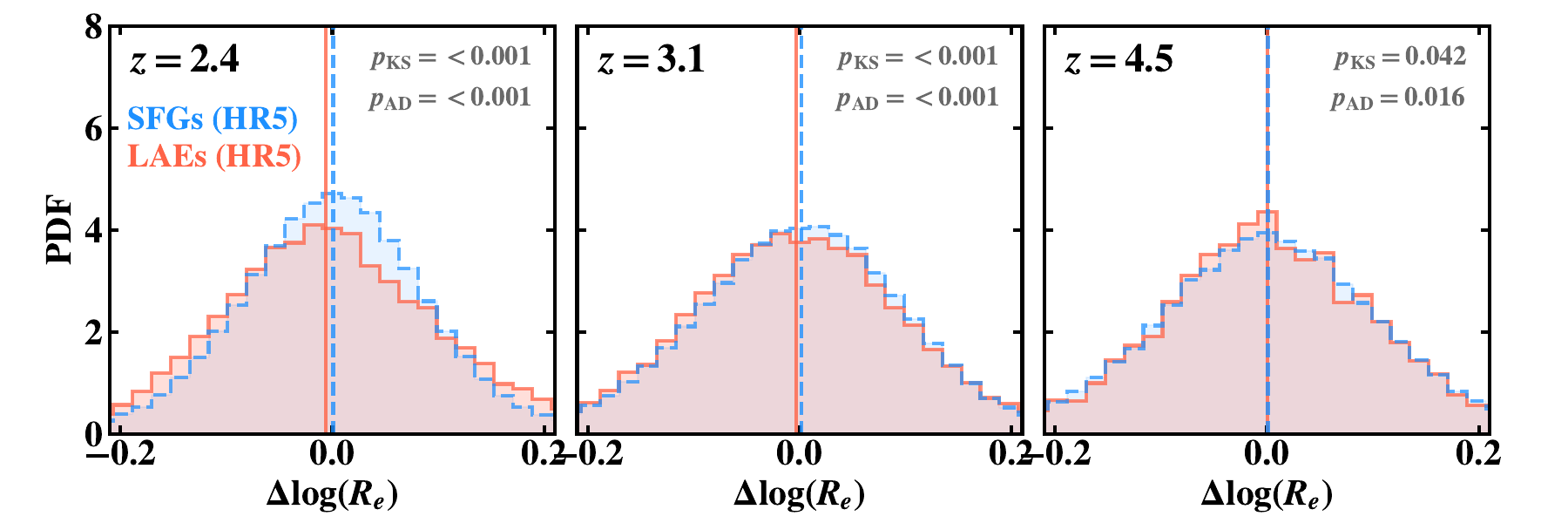}
    \caption{Histogram of the size difference ($\Delta \log R_e$) between each galaxy and the best-fit size-mass relations in the Horizon Run 5 simulation. The meanings of the vertical lines and the texts in each panel are similar to those of Figure \ref{fig:logD_hists}.}
    \label{fig:logD_hists_inHR5}
\end{figure*}

\section{Discussions} \label{sec:Discussions}

\subsection{Comparison with the Horizon Run 5 Simulation} \label{subsec:size-mass_inHR5}

In Section \ref{subsec:rest-optical_morphologies}, we find that LAEs tend to have smaller sizes and larger Sérsic indices than typical SFGs at $z=2.4$ and $3.1$, but those at $z=4.5$ only show a hint of size difference from typical SFGs. Here, we use the data of Horizon Run 5 \citep[HR5;][]{LeeJH_2021_ApJ} cosmological hydrodynamical simulation to examine whether this observed trend could be reproduced. We start from the galaxy catalogs at three snapshots of $z=2.4$, $3.1$, and $4.5$ within a cuboid region of $90 \times 90 \times 1,000 \; \mathrm{(cMpc)}^3$, which has the highest resolution ($\sim1$ physical kpc) in the simulation. To consider only those galaxies with statistically reliable physical properties, we select the galaxies with stellar masses larger than $2.137\times10^8\;M_{\odot}$, which corresponds to a mass of $\sim100$ stellar particles. We use these galaxies to construct samples of typical SFGs by applying a simple specific star-formation rate (sSFR) cut of $\log(\rm{sSFR}/\rm{yr^{-1}}) > -9.5$. We note that using different values for the criterion does not change the main results. The resulting numbers of selected typical SFGs are $318422$, $261468$, and $87504$, for $z=2.4$, $3.1$, and $4.5$, respectively.

\setlength{\arrayrulewidth}{0.6pt}
\begin{deluxetable*}{lccccccccccc}[t!]
    \tabletypesize{\scriptsize}
    \renewcommand{\arraystretch}{1.7}
    \renewcommand{\tabcolsep}{2.0mm}
    \tablecaption{Best-fit parameters of the size-mass relations for our samples using Equation (\ref{eq:size-mass-formula}).}
    \label{table:best-fit_valse}
    \tablehead{\colhead{} & \multicolumn{3}{c}{typical SFGs (COSMOS2025)} & \colhead{} & \multicolumn{3}{c}{ODIN LAEs} & & \multicolumn{3}{c}{typical SFGs (HR5)} \\ \cline{2-4} \cline{6-8} \cline{10-12}
    \colhead{Redshift} & \colhead{$A$ [kpc]} & \colhead{$\beta$} & \colhead{$\log\sigma_{\mathrm{int}}$} & \colhead{} & \colhead{$A $ [kpc]} & \colhead{$\beta$\tablenotemark{a}} & \colhead{$\log\sigma_{\mathrm{int}}$} & & \colhead{$A$ [kpc]} & \colhead{$\beta$} & \colhead{$\log\sigma_{\mathrm{int}}$}} 
    \startdata
    $2.4$ & $2.74^{+0.03}_{-0.03}$ & $0.188^{+0.003}_{-0.003}$ & $0.208^{+0.001}_{-0.002}$ & & $1.99^{+0.15}_{-0.16}$ & $0.188$ & $0.256^{+0.023}_{-0.026}$ & & $5.15^{+0.009}_{-0.009}$ & $0.175^{+0.0004}_{-0.0004}$ & $0.094^{+0.0001}_{-0.0001}$ \\
    $3.1$ & $2.19^{+0.03}_{-0.03}$ & $0.177^{+0.004}_{-0.004}$ & $0.196^{+0.002}_{-0.002}$ & & $1.62^{+0.11}_{-0.13}$ & $0.177$ & $0.247^{+0.021}_{-0.025}$ & & $4.00^{+0.009}_{-0.009}$ & $0.143^{+0.0006}_{-0.0006}$ & $0.104^{+0.0002}_{-0.0002}$ \\
    $4.5$ & $1.48^{+0.04}_{-0.05}$ & $0.097^{+0.009}_{-0.009}$ & $0.205^{+0.003}_{-0.003}$ & & $1.24^{+0.09}_{-0.10}$ & $0.097$ & $0.265^{+0.023}_{-0.027}$ & & $2.82^{+0.015}_{-0.015}$ & $0.076^{+0.0013}_{-0.0013}$ & $0.109^{+0.0003}_{-0.0003}$ \\
    \hline
    \enddata
    \tablenotetext{a}{The slope of the size-mass relation for ODIN LAEs is fixed to be the same as that of typical SFGs at similar redshifts from COSMOS2025 catalog.}
\end{deluxetable*}

For LAEs, we adopt the samples prepared in \cite{Im_2024_ApJ}, and briefly introduce the selection procedure as follows. We first assign a \Lya rest-frame equivalent width (REW) to each galaxy based on the empirical relation between the UV luminosity and the \Lya REW from \cite{Mason_2018_ApJ}. This approach is motivated by empirical studies assuming that LAEs are a subset of Lyman-break galaxies, and that their \Lya REW distribution can be described as a conditional probability function at a given UV luminosity \citep[see e.g., ][]{Dijkstra_Wyithe_2012, Weinberger_2019}. While this empirical modeling cannot determine the exact physical properties of individual LAEs, it allows us to study the statistical properties of the LAE population. We also determine \Lya emission line luminosity $(L_{\rm{Ly\alpha}})$ using the \Lya REW and the UV continuum level, which is calculated from stellar population synthesis modeling. Then we select LAEs by adopting the detection limit of the ODIN NB observations as the minimum of $L_{\rm{Ly\alpha}}$. We also require LAEs to have a \Lya REW larger than $20\;\mathrm{\AA}$, which is the desired minimum \Lya REW of the ODIN LAE-selections (see \citealt{Firestone_2024_ApJ_LAEselection}). A more detailed description on LAE selection in HR5 data is given in Section 2.2 of \cite{Im_2024_ApJ}. The numbers of LAEs at $z=2.4$, $3.1$, and $4.5$ are $12715$, $19552$, and $7116$, respectively.

We measure the $R_e$ and Sérsic index of each HR5 galaxy from the distributions of the stellar particles projected on the $x-y$ plane of the simulation box. First, we determine the major and minor axes using eigenvectors and eigenvalues of the moment of inertia tensor for the stellar particles following \cite{Park_2022}. Then, we construct the $i$-band ($\sim7500\;\mathrm{\AA}$) surface brightness profile as a function of the radius along the major axis. Finally, we fit this profile with a Sérsic model to find best-fit parameters. We exclude galaxies whose uncertainties in the size and Sérsic index are larger than the corresponding best-fit values. This step removes 7426, 6811, and 3141 typical SFGs at $z=2.4$, $3.1$, and $4.5$, respectively (corresponding to $\lesssim3.6\%$ of the parent sample). For LAEs, the numbers of excluded objects are 147, 289, and 170 at $z=2.4$, $3.1$, and $4.5$, respectively (corresponding to $\lesssim2.5\%$ of the parent sample).

Figure \ref{fig:size-mass_inHR5} shows the resulting size-mass relations of LAEs and typical SFGs in HR5. We also present the best-fit power-law relations for HR5 SFGs obtained with Equation (\ref{eq:size-mass-formula}), along with the observational relations derived for typical SFGs from the COSMOS2025 catalog. Here, we assume that the uncertainty in both size and stellar mass is uniform for all galaxies in the simulation. The best-fit parameters are available in Table \ref{table:best-fit_valse}. Although the best-fit normalization ($A$) of the HR5 relations differs from that of the observations (i.e., COSMOS2025), the observed flattening of the size-mass relation for SFGs at higher redshifts is reproduced. The discrepancy in the normalization can be attributed to the dust attenuation effects which are not accounted for in the HR5 galaxies. We also note that the intrinsic scatter in the size-mass relation is smaller in HR5 than in COSMOS2025, likely due to potential interlopers or measurement uncertainties in the observations.

We also calculate the size difference ($\Delta \log R_e$) between the best-fit relation and the size of each galaxy in HR5 using Equation (\ref{eq:logD_formula}). Figure \ref{fig:logD_hists_inHR5} shows the histogram of $\Delta \log R_e$ for LAEs and typical SFGs in the HR5 simulation, along with the $p$-values from the KS- and AD tests. Consistent with the observational results in Section \ref{subsec:size-mass_relations}, LAEs in the simulation tend to have smaller sizes than typical SFGs at $z=2.4$ and $3.1$, while there is only a small difference between them at $z=4.5$. However, the size offset is much smaller than that of observational results (see Figure \ref{fig:logD_hists}). This difference could be attributed to observational uncertainties, including potential interlopers, or it suggests that the empirical modeling adopted for LAEs in the simulation does not fully capture the complex relationship between \Lya emission and rest-frame optical morphology.

\begin{figure*}[t!]
    \centering
    \includegraphics[width=\textwidth]{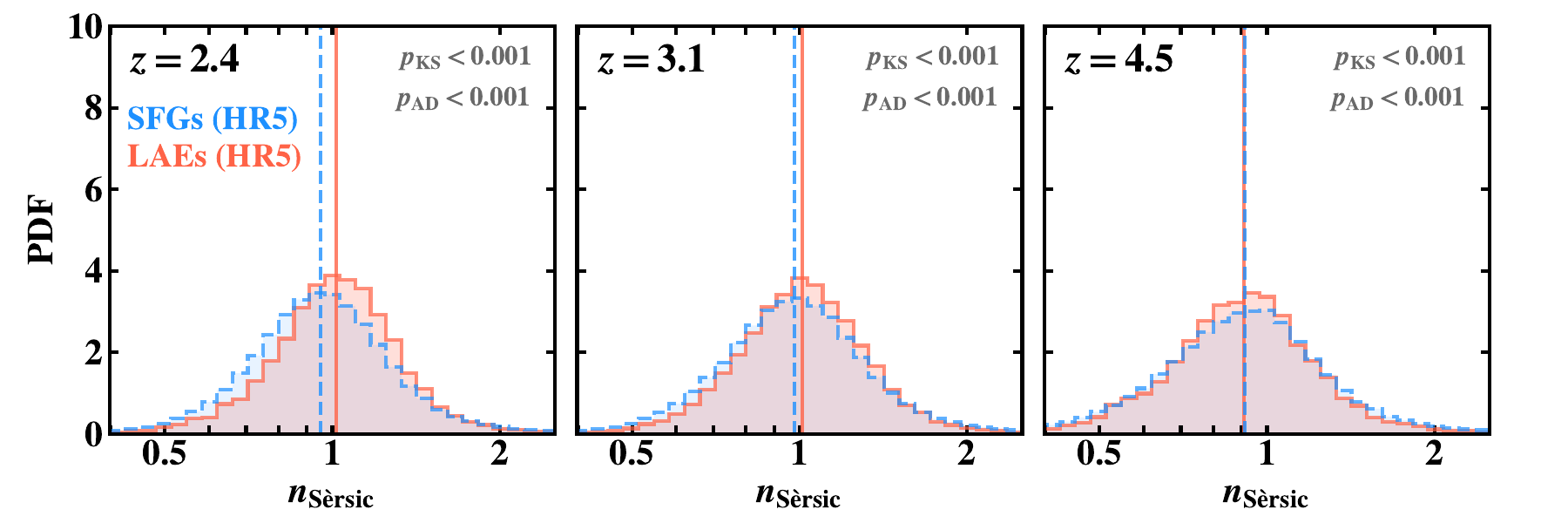}
    \caption{Histogram of the Sérsic index of LAEs and typical SFGs in Horizon Run 5 simulation. The meanings of the vertical lines and the texts in each panel are similar to those of Figure \ref{fig:logD_hists}.}
    \label{fig:nSersic_hists_in_HR5}
\end{figure*}

In Figure \ref{fig:nSersic_hists_in_HR5}, we present the distributions of Sérsic indices for LAEs and typical SFGs in the HR5 simulation. Similar to the observational results in Section \ref{subsec:sersic_indices}, LAEs tend to have larger Sérsic indices than typical SFGs at $z=2.4$ and $3.1$. However, at $z=4.5$, LAEs and typical SFGs have much smaller differences in their Sérsic index, while the $p$-values from KS- and AD tests still suggest that the two distributions are different. The difference between the median Sérsic index of the two populations is smaller than that of observation (see Figure \ref{fig:nSersic_hists_ODIN_LAEs}). Again, this discrepancy could be attributed to uncertainties and potential interlopers in observations. Another possibility is the limitations of the empirical LAE modeling in the simulation, which may not fully account for the complex \Lya photon escape process.

It is possible to use the HR5 data to further explore the formation and evolution of LAEs and the physical conditions for \Lya photon production and escape. However, such a detailed analysis is beyond the scope of this study and could be a topic of future studies.

\subsection{Distinct Evolution in typical Size of LAE Population compared to SFGs}\label{subsec:LAE_evolusion}

To examine whether the LAE population has a distinct size evolution or growth rate compared to the typical SFG population, we investigate the evolution of their typical sizes as a function of redshift. For this analysis, we perform an MCMC fitting similar to that used for the typical SFGs. The resulting best-fit parameters are available in Table \ref{table:best-fit_valse}. Here, we assume that LAEs follow a size-mass relation with the same power-law slope ($\beta$) as typical SFGs, but with a different normalization factor ($A$) and the intrinsic scatter $\sigma_{\mathrm{int}}$. We note that the resulting best-fit parameters are highly uncertain if we leave all three parameters free. 

\begin{figure*}[ht!]
    \centering
    \includegraphics[width=\textwidth]{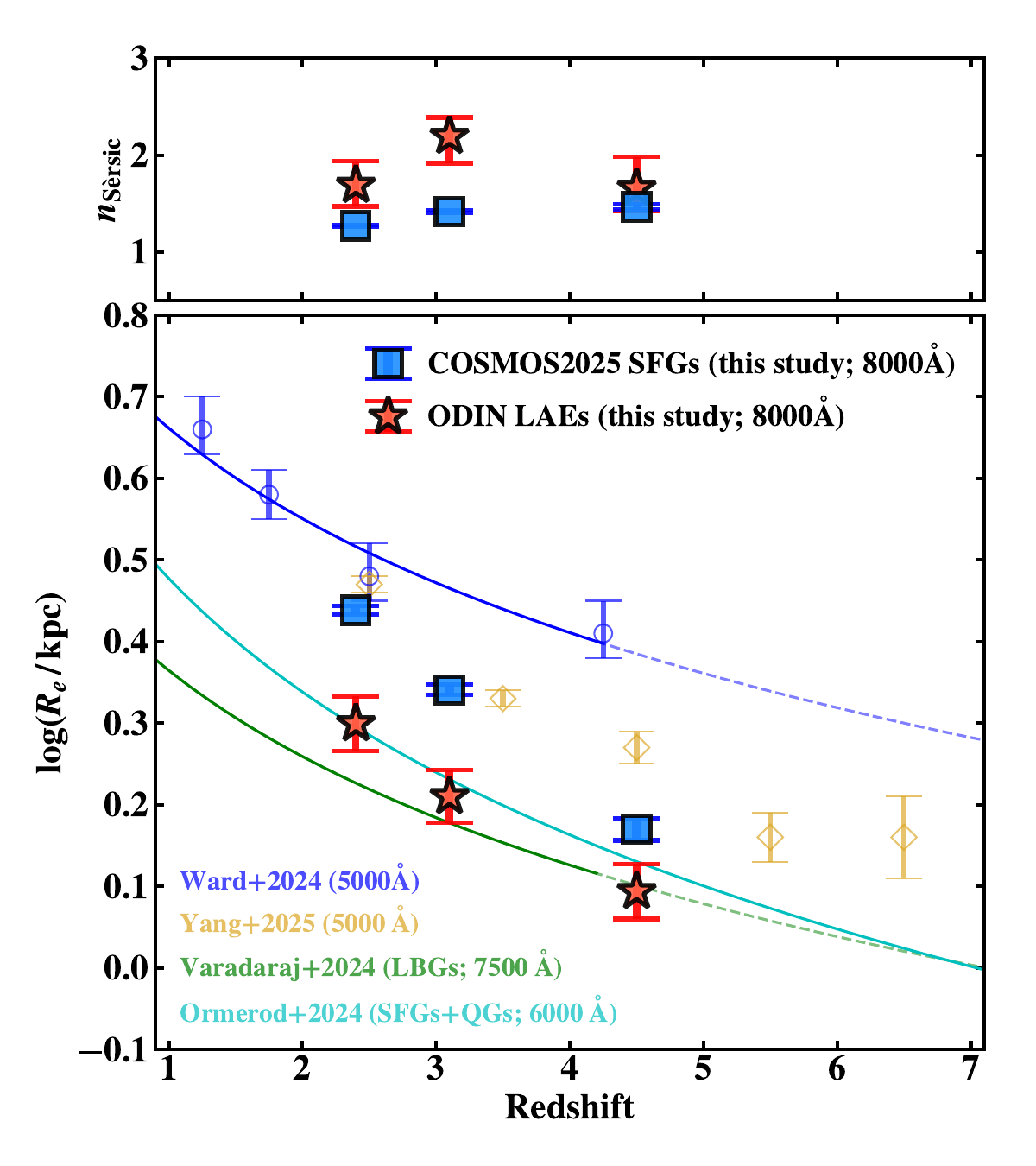}
    \caption{(Upper panel) Redshift evolution of the median Sérsic index for ODIN LAEs and typical SFGs from the COSMOS2025 catalog. The errorbars represent $1\sigma$ uncertainties obtained from bootstrap resampling with 1000 times. (Lower panels) Redshift evolution of the sizes of the samples at a fixed stellar mass of $5\times10^{10}\;M_{\odot}$. The vertical axis represents the best-fit values of the parameter $A$ in Equation (\ref{eq:size-mass-formula}). The errorbars represent $1\sigma$ uncertainties in $A$ derived from the MCMC fittings, and do not include the intrinsic scatter of the samples. For a comparison, we present the results from literature as well, although the wavelengths for their size measurements are slightly different from ours. The dashed lines for some literature indicate an extrapolation beyond the redshift range of the study. We also note that the result of \citet{Varadaraj_2024_MNRAS} represent the size evolution of continuum-selected SFGs, while \citet{Ormerod_2024_MNRAS} used both SFGs and quiescent galaxies for their analysis.}
    \label{fig:LAE_SFG_size_evolutions}
\end{figure*}

Lower panel of Figure \ref{fig:LAE_SFG_size_evolutions} presents the sizes of LAEs and typical SFGs at a fixed stellar mass of $5\times10^{10}\;M_{\odot}$, which corresponds to the normalization $A$ in Equation (\ref{eq:size-mass-formula}). We also present the results from the literature \citep{Ormerod_2024_MNRAS, Ward_2024_ApJ, Varadaraj_2024_MNRAS, Yang_2025_arXiv} in Figure \ref{fig:LAE_SFG_size_evolutions}, even though the rest-frame wavelengths used for their size measurements are slightly different from ours. We find that both LAEs and typical SFGs grow their sizes toward lower redshift, which is consistent with rest-frame UV sizes at fixed UV luminosity \citep[][]{Shibuya_2019_ApJ}. We also observe a trend that the size difference between LAEs and typical SFGs decreases at higher redshifts, which is qualitatively reproduced in the HR5 simulation. This suggests that LAE populations may have size growth rates distinct from typical SFGs, potentially due to differences in their mass accretion rates or the underlying physical mechanisms. This also can suggest that LAEs represent the early stages of galaxy formation experiencing their first starburst, while typical SFGs at the same redshift have generally had more time to grow. Consequently, this evolutionary gap between LAEs and typical SFGs widens at lower redshifts, resulting in the observed distinct size growth rate.

\subsection{Physical Conditions for \Lya Escape}\label{subsec:Lya_escape_conditions}

In Section \ref{subsec:size-mass_relations}, we find that LAEs tend to have smaller rest-frame optical sizes than typical SFGs at similar redshifts. This is consistent with previous studies on rest-frame UV sizes of LAEs using \textit{HST} data \citep[e.g., ][]{Malhotra_2012_ApJ, Kim_2026_ApJ}. There are also several recent studies with rest-frame optical sizes using \textit{JWST}, showing similar results \citep[e.g., ][]{Liu_2024_ApJ, Ning_2024_ApJ, Song2026size}. In this study, we have improved such analyses by providing direct and robust comparisons between LAEs and typical SFGs using large samples with systematically measured sizes and stellar masses in small redshift ranges. 

We also find a negative correlation between the rest-frame optical sizes of ODIN LAEs and their \Lya REW in Section \ref{subsec:REW_correlations}. This is consistent with the studies of rest-frame UV size measurements using \textit{HST} data \citep[e.g., ][]{Bond_2009_ApJ, Bond_2012_ApJ, Paulino-Afonso_2018_MNRAS, Kerutt_2022_A&A}. Because the \Lya escape fraction ($f_{\mathrm{Ly}\alpha}^{\mathrm{esc}}$) is known to be positively correlated with \Lya REW \citep[e.g., ][]{Sobral_2019_A&A, Jones_2025_MNRAS}, this suggests that more compact galaxies tend to have a higher $f_{\mathrm{Ly}\alpha}^{\mathrm{esc}}$, increasing their probability of being observed as LAEs. This implies that small physical size could be a key condition for the efficient escape of \Lya photons. However, a smaller $R_e$ can indicate either a smaller physical size or a higher central concentration of stellar mass. Therefore, the observed negative correlation between $R_e$ and \Lya REW may originate from either or both of these effects. Indeed, in Section \ref{subsec:sersic_indices}, we find that LAEs have larger rest-frame optical Sérsic indices than typical SFGs at $z=2.4$ and $3.1$. Similar to the trend of the size offset between the two populations, the discrepancy in Sérsic indices becomes more pronounced at $z\lesssim3.1$ (see the upper panel of Figure \ref{fig:LAE_SFG_size_evolutions}). The physical origin of this redshift evolution remains to be explored in future studies. We note that we compare the Sérsic indices themselves without correcting for the stellar-mass effect, as they have a weak and uncertain relationship with stellar mass. (see Section \ref{subsec:sersic_indices}). Furthermore, we observe hints of positive correlation between the Sérsic index and \Lya REW, indicating that central concentration could be another key condition for \Lya photon escape. Specifically, a higher concentration may enhance the impact of supernova feedback, efficiently ionizing neutral hydrogen in star-forming clouds. This interpretation aligns with studies with numerical simulations showing that supernova feedback plays a crucial role in \Lya photon escape from galaxies \citep{Kimm_2014, Trebitsch_2017, Kakiichi_2021}.

To determine which factor is dominant, a similar analysis on LAE samples with a fixed Sérsic index (or size) would be necessary, requiring a larger sample size. Furthermore, \Lya REW has strong correlation with stellar mass as well (see Appendix \ref{sec:LyaREW_vs_Mstar_or_logD}); when examining the correlation between \Lya REW and the size excess relative to the size-mass relation ($\Delta \log R_e$; see Equation (\ref{eq:logD_formula})), we find no significant correlation between the two (see Appendix \ref{sec:LyaREW_vs_Mstar_or_logD}). It is also plausible that the primary effect is driven by stellar mass, independently of size, and that the observed correlation is merely a secondary effect. This is consistent with \citet{Weiss_2021}, who found that $f_{\mathrm{Ly}\alpha}^{\mathrm{esc}}$ of $[\mathrm{OIII}]$ emitters at $1.9<z<2.35$ is more tightly correlated with stellar mass than with the effective radius measured from \textit{HST} images. A larger sample of LAEs is required to distinguish the effect of size, independent of stellar mass.

On the other hand, in Section \ref{subsec:SFMS}, we find that LAEs tend to be located above the star-forming main sequence line, indicating that they are preferentially in starbursting phase with higher starburstiness. This is consistent with the results of recent \textit{JWST} observations \citep[e.g.,][]{Liu_2024_ApJ, Ning_2024_ApJ} and cosmological simulations \citep[e.g.,][]{Im_2024_ApJ}. This result suggests that extreme star-formation would be another key physical condition for a galaxy to become an LAE \citep[see e.g., ][]{Matthee_2021_MNRAS, Kim_2026_ApJ}. Furthermore, in Section \ref{subsec:REW_correlations}, we find a strong positive correlation between \Lya REW and the ratio of the instantaneous star-formation rate to that averaged over the last $100\;\mathrm{Myr}$ (i.e., $\mathrm{SFR_{inst}}/\mathrm{SFR_{100 Myr}}$). This is consistent with a recent ODIN result, showing that most of the ODIN LAEs are experiencing their most significant starburst in their star formation history \citep{Firestone_2025_ApJ_LAE_SFHs}. As mentioned in Section \ref{subsec:REW_correlations}, a strong positive correlation between \Lya REW and $\mathrm{SFR_{inst}}/\mathrm{SFR_{100 Myr}}$ is  expected. Therefore, the observed correlation indicates that the SED fitting of the COSMOS2025 catalog successfully reconstructs the recent SFH, likely due to the medium-band photometry. Furthermore, this implies that the complex \Lya radiative transfer process does not significantly distort or erase the intrinsic relation between \Lya REW and $\mathrm{SFR_{inst}}/\mathrm{SFR_{100 Myr}}$.

\section{Conclusions}\label{sec:Conclusions}

We analyzed rest-frame optical ($\sim8000\mathrm{\AA}$) morphologies and star formation activities of \Lya emitters (LAEs) at $z=2.4$, $3.1$, and $4.5$ in comparison with typical SFGs at similar redshifts. We utilized the \textit{JWST}/NIRCam $\mathrm{F277W}$ and $\mathrm{F444W}$ images of the COSMOS-Web survey to examine their rest-frame optical morphologies. We also adopted the physical parameters of stellar mass and star formation rates derived from the \texttt{CIGALE} SED fitting of the COSMOS2025 catalog. Our primary results are as follows.

\begin{enumerate}

    \item [(I)] We compared the size-mass relations of LAEs and typical SFGs, and found that LAEs tend to have smaller sizes than typical SFGs across all three redshifts. We found that LAEs have larger Sérsic indices than typical SFGs at $z=2.4$ and $3.1$, whereas we could not find statistically significant difference at $z=4.5$. We also examined the star-formation main-sequence relations and found that LAEs are located above the main sequence. These findings show the compact and starbursting nature of LAEs compared to SFGs, which are also reproduced in the Horizon Run 5 cosmological hydrodynamical simulation. 
    
    \item [(II)] We found that the rest-frame equivalent width (REW) of the \Lya emission line has negative and positive correlations with the rest-frame optical size and the Sérsic index, respectively. Given the known positive correlation between the \Lya escape fraction ($f_{\mathrm{Ly}\alpha}^{\mathrm{esc}}$) and the \Lya REW, this finding suggests that galaxies with smaller physical sizes and higher central concentrations favor higher $f_{\mathrm{Ly}\alpha}^{\mathrm{esc}}$, increasing their probability of being observed as LAEs.
    
    \item [(III)] We found strong positive correlations between the Lya REW and the ratio of the current star-formation rate to that averaged over the last $100\;\mathrm{Myr}$ (i.e., $\mathrm{SFR_{inst}}/\mathrm{SFR_{100 Myr}}$), which is expected. This indicates that the SED fitting of the COSMOS2025 catalog successfully reconstructs recent SFH, and that the \Lya radiative transfer process does not significantly distort the intrinsic relationship.
    
    \item [(IV)] We found that the size difference between LAEs and typical SFGs decreases with increasing redshift. This trend results in a different slope for the size evolution of the LAE population compared to typical SFGs. This finding indicates that the LAE population has a distinct growth rate of their typical sizes compared to that of SFGs, implying differences in their mass accretion mechanisms or rates.

\end{enumerate}

Overall, most of our findings are consistent with the previous observational results on rest-frame UV size measurements using \textit{HST} data. However, thanks to \textit{JWST} and large survey programs such as ODIN and COSMOS-Web, we were able to extend these analyses into the rest-frame optical and make a more robust comparison between LAEs and typical SFGs. We also emphasize that these findings are qualitatively consistent with the predictions from the HR5 cosmological hydrodynamical simulation. The HR5 simulation data can offer a valuable opportunity to investigate the physical origins of these distinct LAE features in more detail. Furthermore, once the ODIN survey is completed, we anticipate having a much larger sample of LAEs. When combined with future large or deep \textit{JWST} imaging surveys, this will substantially improve the robustness of these types of analyses, providing a more detailed picture of LAE morphologies and the physical mechanisms responsible for the \Lya photon escape.

\begin{acknowledgments}
We thank the reviewer for the insightful comments. HSH acknowledges support from the National Research Foundation of Korea (NRF) funded by the Korea government (MSIT; RS-2026-25482692) and the Global-LAMP Program funded by the Ministry of Education (RS-2023-00301976). J.H.L. acknowledges support from the Basic Science Research Program through the National Research Foundation of Korea (NRF) funded by the Ministry of Education (No. RS-2024-00452816). LG gratefully acknowledges support from the FONDECYT regular project number 1230591, the ANID BASAL project FB210003, ANID - MILENIO - NCN2024\_112. CBP is supported by the KIAS Individual Grant PG016904 at the Korea Institute for Advanced Study (KIAS) and by the National Research Foundation of Korea (NRF) grant funded by the Korean government (MSIT; RS-2024-00360385). H. Song was supported by the National Research Foundation of Korea(NRF) grant funded by the Korea government (MSIT) (No. RS-2025-25442707). This work is supported by the Center for Advanced Computation at Korea Institute for Advanced Study. This work is partially supported by the grants GALBAR ANR-25-CE31-4684 of the French Agence Nationale de la Recherche. This project has received funding from the European Union’s Horizon 2020 research and innovation programme under the Marie Skłodowska-Curie grant agreement No 101148925. This work is based on observations made with the NASA/ESA/CSA James Webb Space Telescope. The data were obtained from the Mikulski Archive for Space Telescopes (MAST) at the Space Telescope Science Institute, which is operated by the Association of Universities for Research in Astronomy, Inc., under NASA contract NAS 5-03127 for JWST. These observations are associated with program \#1727 (COSMOS-Web). The data analyzed here can be obtained from the MAST archive at \dataset[10.17909/4969-s888] {https://doi.org/10.17909/4969-s888}.
\end{acknowledgments}

%


\appendix

\begin{figure}[ht!]
    \centering
    \includegraphics[width=\linewidth]{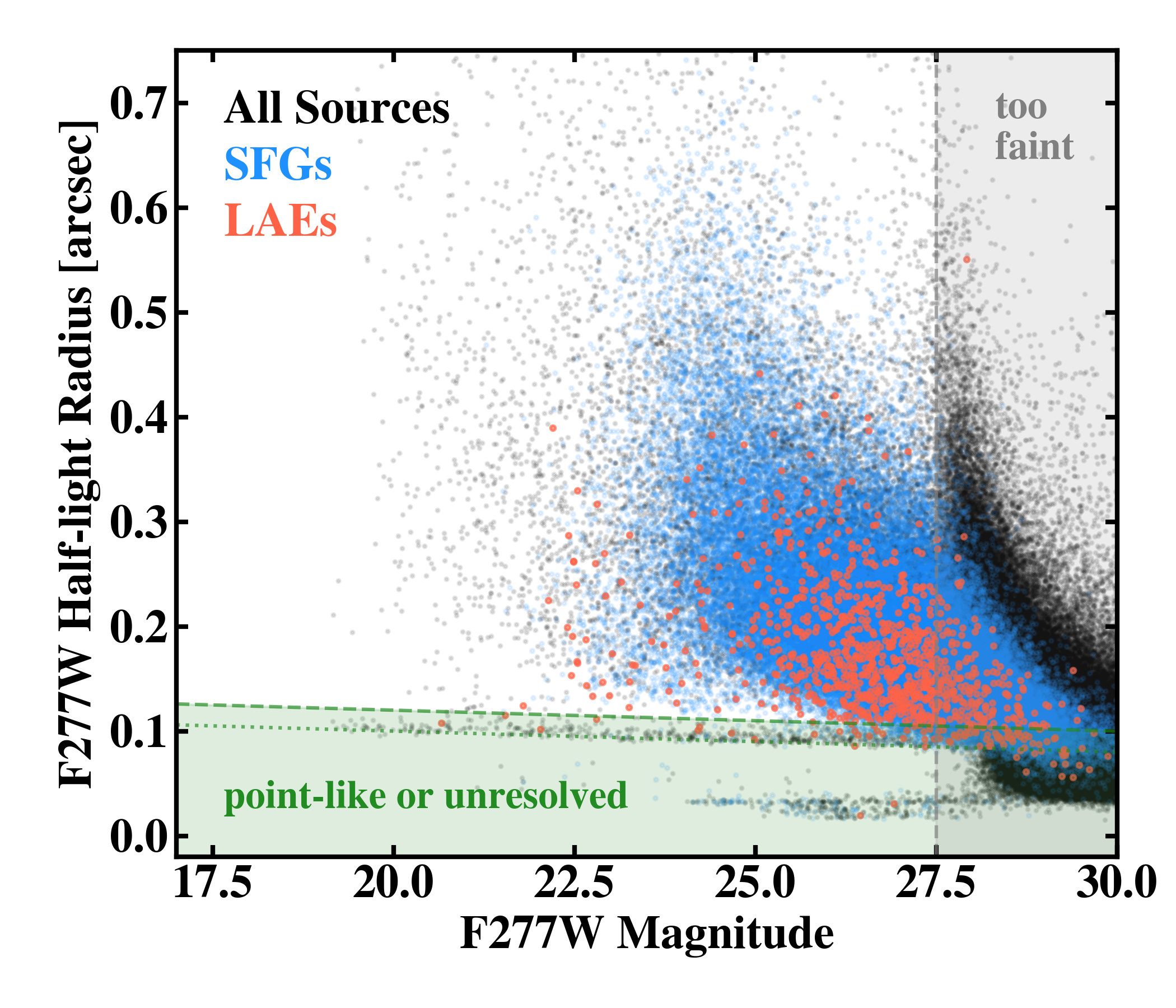}
    \caption{Relation between the half-light radius ($r_h$) and the magnitude in F277W. The green dotted line denotes the location of the tight relation for the point sources. We exclude the sources with $r_h$ similar to or smaller than that of the point sources (i.e., below the green dashed line) from our sample. The gray shaded region indicates the magnitude cut for faint sources ($> 27.5$; see Section \ref{subsec:COSMOS-Web_images})}
    \label{fig:rh_vs_mag}
\end{figure}

\begin{figure*}[ht!]
    \centering
    \includegraphics[width=\textwidth]{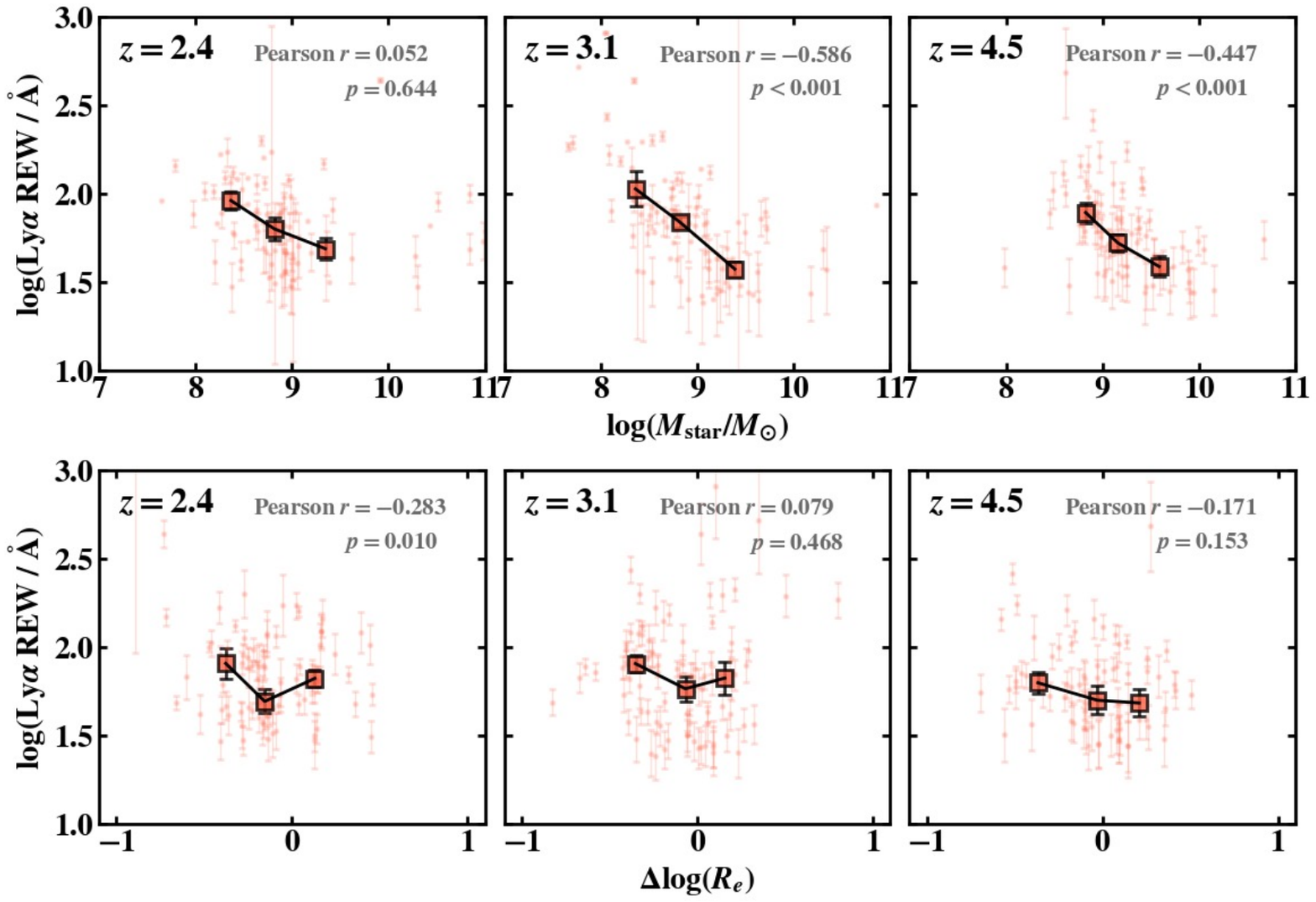}
    \caption{Relation of \Lya REW with stellar mass (upper panels), and with the size excess relative to the size-mass relation (lower panels) for ODIN LAEs at $z=2.4$, $3.1$, and $4.5$. While \Lya REW has strong correlations with stellar mass, there are only weak relations between the \Lya REW and $\Delta \log R_e$ at $z=2.4$ and $4.5$.}
    \label{fig:appendix}
\end{figure*}

\section{Point Source Rejection} \label{sec:point_source_rejection}

Point sources are known to have a tight correlation between the half-light radius ($r_h$) and the magnitude \citep[see e.g., ][]{Bertin_2011_ASPC, Skelton_2014}. To identify point sources, we examine the relation between $r_h$ and the magnitude in F277W images, measured by the \texttt{SExtractor}. Figure \ref{fig:rh_vs_mag} shows the relation for our sample galaxies, along with that of all sources identified by \texttt{SExtractor}. We find a tight correlation of point sources (the green dotted line in the Figure), as in similar studies \citep[e.g., ][]{Holwerda_2024, Ward_2024_ApJ}. We classify the sources with $r_h$ smaller than $0\farcs02+ r_h$ of point sources (the green dashed line) as point-like or unresolved sources, and exclude them from our sample. For visualization, we also present the region where we remove too faint sources (F277W magnitude $ > 27.5$; see Section \ref{subsec:galfit}) i.e., we use only the sample galaxies that fall on the upper left region of Figure \ref{fig:rh_vs_mag}.

\section{\Lya REW dependence on and $M_{\mathrm{star}}$ and the $\Delta \log R_e$} \label{sec:LyaREW_vs_Mstar_or_logD}

We present the relationship between \Lya REW and stellar mass, and with the size excess relative to the size-mass relation ($\Delta \log R_e$; defined by Equation (\ref{eq:logD_formula})), in Figure \ref{fig:appendix}. We find that \Lya REW has a strong correlation with stellar mass, whereas it has only a weak correlation with $\Delta \log R_e$. Larger samples of LAEs are necessary to decouple the effect of galaxy size on \Lya REW (i.e., on \Lya escape fraction; $f_{\mathrm{Ly}\alpha}^{\mathrm{esc}}$) from that of stellar mass.


\bibliography{refs}{}
\bibliographystyle{aasjournalv7}



\end{document}